\documentclass[twoside,11pt]{article}

\usepackage[accepted]{melba}

\usepackage{amsmath,amsfonts}

\usepackage{booktabs}

\usepackage{color}
\usepackage{hyperref}
\usepackage{marginnote}

\long\def\dis#1{\vskip 0.3in\noindent{\large\bf Disclaimer}\vskip 0.2in \noindent #1}

\melbaheading{2021:018}{https://www.melba-journal.org/papers/2021:018.html}{2021}{1-28}{06/2021}{11/2021}{Gottschalk, Maier, Kordon and Kreher}{}{}

\ShortHeadings{View-consistent metal segmentation}{Gottschalk, Maier, Kordon and Kreher}
\firstpageno{1}

\title{View-Consistent Metal Segmentation in the Projection Domain for Metal Artifact Reduction in CBCT – An Investigation of Potential Improvement}

\author{\name T.M.~Gottschalk \email tristan.gottschalk@fau.de \\
    \addr Pattern Recognition Lab, Department of Computer Science, Friedrich-Alexander University Erlangen-Nuremberg, Erlangen 91058, Germany\\
    Erlangen Graduate School in Advanced Optical Technologies (SAOT), Friedrich-Alexander University Erlangen-Nuremberg, Erlangen 91052, Germany \\
    Siemens Healthcare GmbH, Forchheim 91301, Germany \\
    \AND
    \name A.~Maier \email andreas.maier@fau.de\\
    \addr Pattern Recognition Lab, Department of Computer Science, Friedrich-Alexander University Erlangen-Nuremberg, Erlangen 91058, Germany\\
    Machine Intelligence, Department of Computer Science, Friedrich-Alexander University Erlangen-Nuremberg, Erlangen 91058, Germany\\ 
    Erlangen Graduate School in Advanced Optical Technologies (SAOT), Friedrich-Alexander University Erlangen-Nuremberg, Erlangen 91052, Germany \\
    \AND
    \name F.~Kordon \email florian.kordon@fau.de \\ 
    \addr Pattern Recognition Lab, Department of Computer Science, Friedrich-Alexander University Erlangen-Nuremberg, Erlangen 91058, Germany\\
    Erlangen Graduate School in Advanced Optical Technologies (SAOT), Friedrich-Alexander University Erlangen-Nuremberg, Erlangen 91052, Germany \\
    Siemens Healthcare GmbH, Forchheim 91301, Germany \\
    \AND
    \name B.W.~Kreher \email bjoern.kreher@siemens-healthineers.com\\
    \addr Siemens Healthcare GmbH, Siemensstraße 1, Forchheim 91301, Germany \\
}

\begin{document}

\maketitle

\begin{abstract}
The positive outcome of a trauma intervention depends on an intraoperative evaluation of inserted metallic implants. Due to occurring metal artifacts, the quality of this evaluation heavily depends on the performance of so-called Metal Artifact Reduction methods (MAR). The majority of these MAR methods require prior segmentation of the inserted metal objects. Therefore, typically a rather simple thresholding-based segmentation method in the reconstructed 3D volume is applied, despite some major disadvantages. With this publication, the potential of shifting the segmentation task to a learning-based, view-consistent 2D projection-based method on the downstream MAR's outcome is investigated. For segmenting the present metal, a rather simple learning-based 2D projection-wise segmentation network that is trained using real data acquired during cadaver studies, is examined. To overcome the disadvantages that come along with a 2D projection-wise segmentation, a Consistency Filter is proposed. The influence of the shifted segmentation domain is investigated by comparing the results of the standard fsMAR with a modified fsMAR version using the new segmentation masks. With a quantitative and qualitative evaluation on real cadaver data, the investigated approach showed an increased MAR performance and a high insensitivity against metal artifacts. For cases with metal outside the reconstruction's FoV or cases with vanishing metal, a significant reduction in artifacts could be shown. Thus, increases of up to roughly 3 dB w.r.t. the mean PSNR metric over all slices and up to 9 dB for single slices were achieved. The shown results reveal a beneficial influence of the shift to a 2D-based segmentation method on real data for downstream use with a MAR method, like the fsMAR. The nature of the method further suggests the same beneficial behavior for all (also recent data-driven) MAR methods, that for now comprise a 3D-volume-based segmentation step for subsequent inpainting.
%
\end{abstract}


\section{Introduction}
\label{sec:introduction}
\begin{figure}[!tbp]
\centering
\includegraphics[scale=0.9]{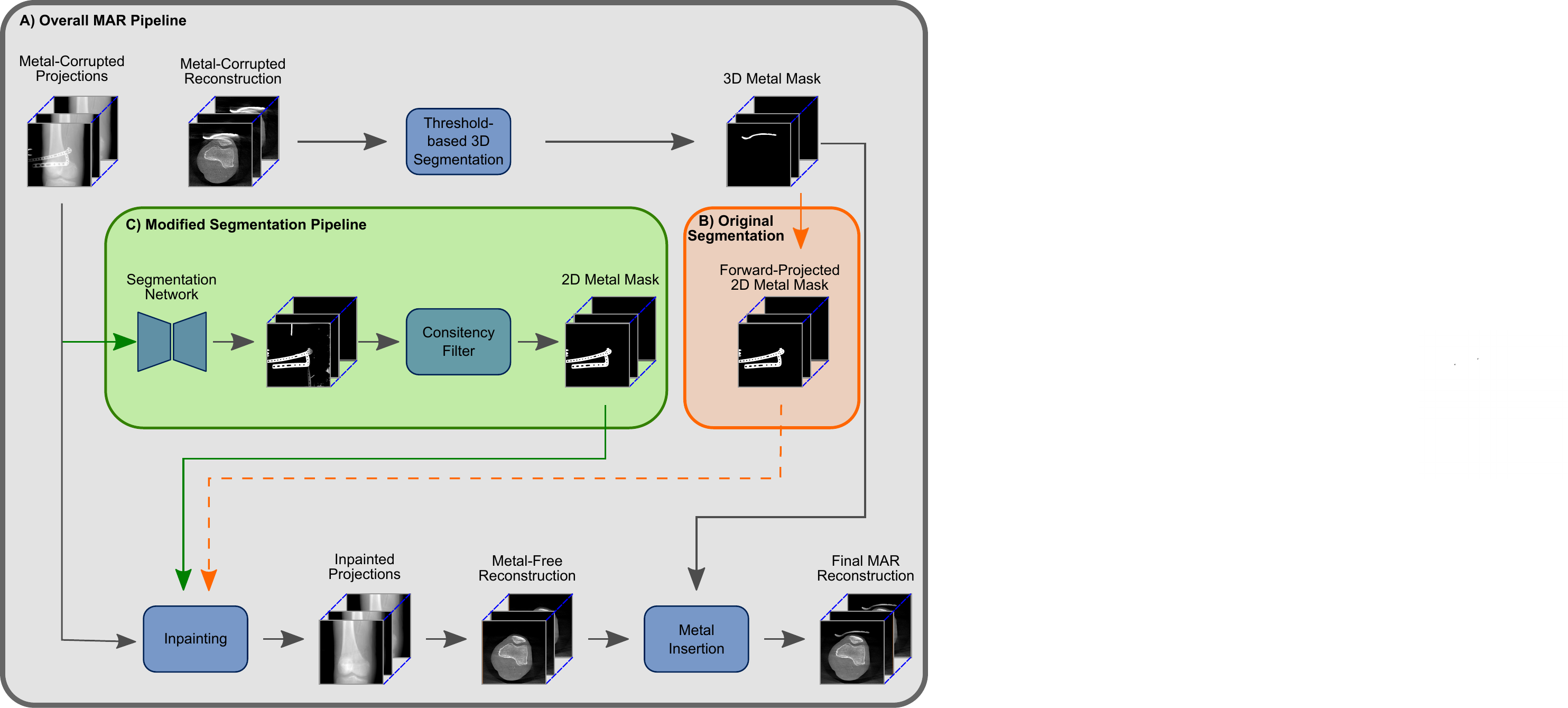}
\caption{Illustration of the simplified processing steps of an inpainting-based MAR method as used by e.g. the fsMAR. Further, the modifications done to the segmentation steps using the proposed method are shown. Whereas A) represents the overall MAR pipline setting the basis for both variants, B) represents the segmentation steps done when using a threshold-based 3D segmentation with subsequent forward-projection of the extracted masks to 2D, C) illustrates the segmentation step of the proposed modified fsMAR. It consists of a data-driven 2D segmentation with a subsequent Consistency Filter. The dashed orange arrow originating from B) denotes the connection used by the standard fsMAR and which is then replaced by the connection originating from C) when using the proposed segmentation method.}
\label{fig:basic_mar}
\end{figure}

In the context of trauma and orthopedic interventions, the intraoperative evaluation of a successful fracture reduction and the correct positioning of the inserted metal implants are crucial for the outcome of the intervention and the patient's healing process. For that evaluation, in many cases, a 3D reconstruction is vital. With current generations of movable C-arm systems, like the Siemens Cios Spin\textsuperscript{\textcopyright}, this is possible inside the operation room with a limited amount of additional effort. However, due to so-called metal artifacts and the corresponding lack of image quality, even these 3D reconstructions do not allow for a profound evaluation. Consequently, there exists the necessity of well-functioning Metal Artifact Reduction methods (MAR). 
Alongside more traditional MAR methods \citep{Meyer2010,Xinhui2008} like e.g. the \textit{frequency split metal artifact reduction} (fsMAR) \citep{Meyer2012}, in recent years, many data-driven MAR approaches were developed \citep{Claus2017,Ghani2018,Gjesteby2017,Huang2018,Park2018,ketcha2021}. Besides the majority of the more recent MAR approaches, the fsMAR is an inpainting-based MAR, which typically consists of three major steps as shown in Fig.~\ref{fig:basic_mar}, (1) segmentation of the present metal objects in the initial unprocessed 3D reconstruction (thus containing metal artifacts), (2) 2D projection/sinogram-based inpainting in the previously segmented metal regions and (3) a so-called metal insertion in the subsequently reconstructed artifact-reduced 3D volume. Consequently, the MAR's outcome heavily depends on the quality of the performed metal segmentation. Since non-segmented metal will not be processed by the downstream steps of the MAR method at all, they are still causing artifacts in the final processed reconstruction (cf. Fig.~\ref{fig:res_1}, TS 3, fsMAR). Falsely segmented anatomical structures, however, will potentially cause blurry or completely vanished representations of the very same. Nonetheless, despite its importance \citep{Stille2013,Yu2021}, the segmentation process itself and its influence on the different MAR's outcomes is an underrepresented research topic.

The majority of the mentioned MAR methods solve the segmentation task by a 3D volume-based thresholding mechanism, whose 3D mask is then forward-projected to the projection/sinogram-domain for downstream inpainting (cf. Fig.~\ref{fig:basic_mar}, B)) , although that has two clear disadvantages. First, the metal segmentation is done on the initial and thus metal-artifact-corrupted 3D reconstruction. This severely aggravates the segmentation task or even makes it completely impossible, e.g. for cases with vanishing intensity and contrast due to e.g. photon starvation or scatter (cf. failed mask in Fig.~\ref{fig:3d_seg_compare}, c)). The second disadvantage of the performed 3D-volume-based metal segmentation is caused by the fact that only those metal objects can be detected, that lie inside the field-of-view (FoV) of the corresponding reconstruction. However, the undetectable metal outside the FoV can still cause artifacts spreading across the whole reconstruction.
In trauma interventions, where a high amount of metallic surgical instruments/tools and K-wires exceeding the FoV are used, this is a common case. 

As an alternative to a segmentation of the metal in the 3D reconstructed volume, there exists the possibility of solving that task on the 2D projection images. However, 2D projection-based approaches need to cope with a lower signal-to-noise ratio (SNR). Further, a threshold for the segmentation of metal is not well defined in the projection domain. This is due to the pixel intensity being proportional to the integral of the different materials along the X-ray path. This makes a threshold-based approach impractical. Furthermore, 2D segmentation approaches that work projection-wise cannot account for the necessary consistency of the segmentation throughout the complete 3D scan's set of projections. The resulting inconsistencies in the segmentation lead to so-called secondary artifacts, which themselves show as disruptive streaks in the final MAR's reconstructed volume \citep{barrett2004}. Nonetheless, in contrast to a 3D approach, it holds the advantages of being able to segment all present metal objects (even those lying outside the reconstruction's FoV) and that it does not get deteriorated from metal artifacts like e.g. photon starvation, which are present in the corresponding 3D reconstructions.

Moreover, in recent years data-driven approaches reached impressive results in natural image segmentation \citep{badrinarayanan2017,Chen2018,Long2015}, as well as medical image segmentation tasks. In the context of medical images, the \textit{U-Net} proposed by \cite{Ronneberger2015} and other U-Net-based approaches, like the \textit{nnU-Net} published by \cite{isensee2021} had a significant impact.

Considering the mentioned advantages of segmenting even metal outside the FoV of the reconstructions, as well as the insensitivity of the projection images against metal artifacts, we investigate a rather simple data-driven, view-consistent 2D projection-based metal segmentation approach, which was initially presented in \cite{Gottschalk2021}. To cope with the disadvantage of potentially missing consistency of the segmentation masks, we coupled the segmentation network with a subsequent Consistency Filter (CF). Besides the enforcement of the consistency, the major advantage of the CF lies in the simultaneous removal of  false-positive segmentation, as already shown in \cite{Gottschalk2021}. However, the previous publication lacks an evaluation of the benefits of that method in the corresponding MAR's reconstructions. The mentioned evaluation is done with a cone-beam-based implementation of the more classic fsMAR as currently used in the Siemens Cios Spin\textsuperscript{\textcopyright} C-arm system. This is because the more recent approaches \citep{Claus2017,Ghani2018,Gjesteby2017,Huang2018,Park2018, ketcha2021} either does not include an explicit metal segmentation step as \cite{ketcha2021}, or were developed for CT-based acquisitions (fan-beam geometry). Thus they can not easily applied to cone-beam-based CT (CBCT) reconstructions (incomparably more complex Radon transform) as typically used in interventional setups. Considering that this work neither claims to propose a new segmentation approach, nor an overall new MAR method, but rather investigates potential benefits of shifting the segmentation task from 3D to 2D, using a non-data-driven base method for the evaluation is reasonable.

%
%

\section{Methods}
\label{sec:methods}

\subsection{Acquisition of Data}
\label{sec:data}

\begin{figure}[tb]
\centering
\includegraphics[scale=1.2]{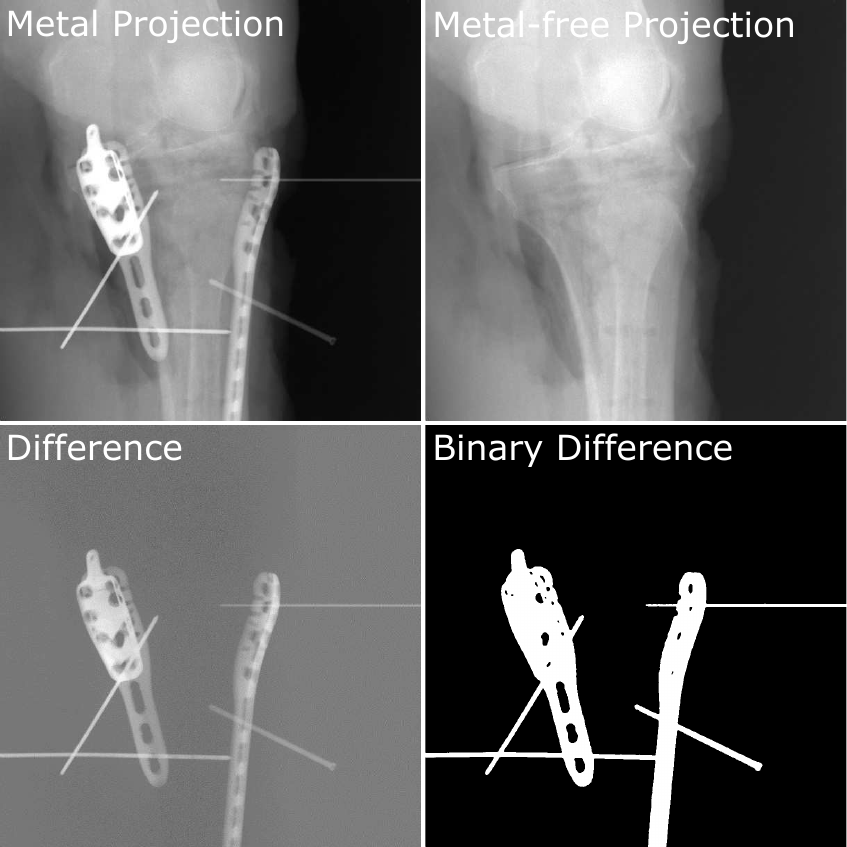}
\caption{The binary GT segmentation mask is generated with the help of the acquired metal-corrupted and metal-free projections. First, the difference image is generated by division. Subsequently, it is binarized.}
\label{fig:label_gen}
\end{figure}

Training a deep learning-based projection-wise metal segmentation in a supervised fashion requires the availability of metal corrupted projection images and their corresponding segmentation labels. For real clinical cases, such segmentation labels need to be created by hand, which oftentimes demands an uneconomically high amount of effort. As a common alternative, corresponding datasets are created using simulation methods, like e.g. Monte Carlo simulations like MC-GPU \citep{Badal2009} or the DeepDRR framework proposed by \cite{Unberath2018}. However, simulations cannot completely reproduce the underlying complex physical image formation process and are thus still inferior concerning clinical realism. To cope with that, the investigated approach is trained using real data, that was acquired during three cadaver studies of human knees, as well as human spines. To discard the necessity of hand labeling for the cadaver datasets, two consecutive 3D scans were performed during these studies -- the first scan with and the second scan without metal implants. Thus, the corresponding segmentation label can simply be generated by dividing the acquired matching projection images. An example can be seen in Fig.~\ref{fig:label_gen}. The 3D scans were performed using a Siemens Cios Spin\textsuperscript{\textcopyright} C-arm system. Each of the 3D scans consists of 400 projection images acquired using an equiangular increment of 0.5$^{\circ}$, thus covering an angular range of 200$^{\circ}$. The projection images themselves have a size of 976x976 pixels with a pixel size of 0.310 mm. To generate suitable label data, the difference image needs to be motion-free. Therefore, neither the acquisition device nor the specimen is allowed to move in between the two consecutive 3D scans. This was ensured by fixating the specimen with clamps to the acquisition table and by remote-controlled handling of the C-arm device from its attached trolley.
During the three cadaver studies, 55 matching 3D scans (36 knees, 19 spines) with and without metal could be acquired, thus ending up with a total amount of 22.000 matching 2D projection images. These comprise 3D scans with different amounts of present metal implants. Furthermore, the positioning of the metal implants varies from being simply placed on the surface of the skin of the specimen, over being placed directly on and inside the bone in a non-clinical fashion, to metal implants being correctly placed with clinical realism. Consequently, the acquired dataset should roughly cover the variety of surgical interventions, which oftentimes comprise randomly placed metal objects/tools outside the FoV of the actual reconstruction and different amounts of implants, depending on the complexity of the fracture. 
Using these 55 3D scans a 5-fold cross-validation is performed. Therefore, the samples were randomly shuffled and then split into five data chunks of 11 3D scans each. For each fold of the cross-validation, three data chunks are assigned for training, one for validation and one for testing. The chunks are selected using a sliding window, such that each data chunk is once assigned to the test set of a fold. Consequently, all 55 datasets at hand are once part of the performed evaluation.
Alongside the 400 2D projection images for each of the acquired 3D scans, the C-arm system already provides us with two reconstructions for each pair of matching scans - one initial non-MAR-processed reconstruction and one fsMAR-processed reconstruction\footnote{All shown reconstructions in this publication are generated using the currently implemented reconstruction pipeline of the Siemens Cios\textsuperscript{\textcopyright} Spin C-arm system.}. These reconstructed volumes have a size of 512x512x512 voxels with a voxel size of 0.313 mm.

\subsection{Segmentation Framework}
\label{sec:seg_frame}

\subsubsection{Network Architecture}
\label{sec:network_arch_seg}

The trained segmentation network is a slight variation of the proposed network in \cite{Gottschalk2021} and thus is again an adapted U-Net-like network. The network is constructed with five layers of contraction-blocks and five layers of corresponding expansion-blocks. Each related pair of contraction and expansion blocks is connected via skip-connections creating gradient flow shortcuts that help to avoid vanishing gradients during the training process \citep{Drozdzal2016}. The contraction-blocks consist of two 2D-convolutional layers followed by one 2D-max-pooling layer. The expansion-blocks comprise a 2D-upsampling layer followed by two 2D-convolutional layers. All convolutional layers use a 3x3 kernel, rectified linear units (ReLU) as activation function, and a batch normalization is employed. The last convolutional layer of the network, however, uses a linear activation function. The network's first contraction-block generates 32 feature maps and each consecutive block doubles (contraction) or halves (expansion) the number of feature maps. Thus, the network's bottleneck holds the amount of 1024 feature maps. The network is designed such that the output image size matches the input image size.

To train the network, the acquired RAW data is transformed to line integral data by applying Lambert-Beer Law \citep{maier2018}. As explained earlier, the network was trained five times using the different folds of the cross-validation. For every fold, the network was trained from scratch to convergence using a batch-size of 8. Apart from applying a cross-entropy-based loss function, the initial learning rate of $1e^{-4}$ was decreased using an exponential decay and Adam \citep{kingma2014} was used as optimizer. The weights are initialized using He initialization \citep{He2015}. Additionally, an online augmentation scheme, inspired by \cite{Isensee2019}, was applied to fully utilize the benefits of patch-based training and to prevent overfitting. During that augmentation, randomized contrast and brightness scaling, rotations, left-right flips, and different amounts of added Poisson noise (to mimic varying image acquisition qualities) were applied. During the experiments regarding the different patch-sizes, as explained in Sec.\ref{sec:patch}, the used patches were acquired by adding randomized cropping to the online augmentation scheme. All experiments were performed using TensorFlow v1.13 \citep{abadi2016}.

\subsubsection{Consistency Filter}
\label{sec:CF}

\begin{figure}[tb]
\centering
    \includegraphics[scale=0.85]{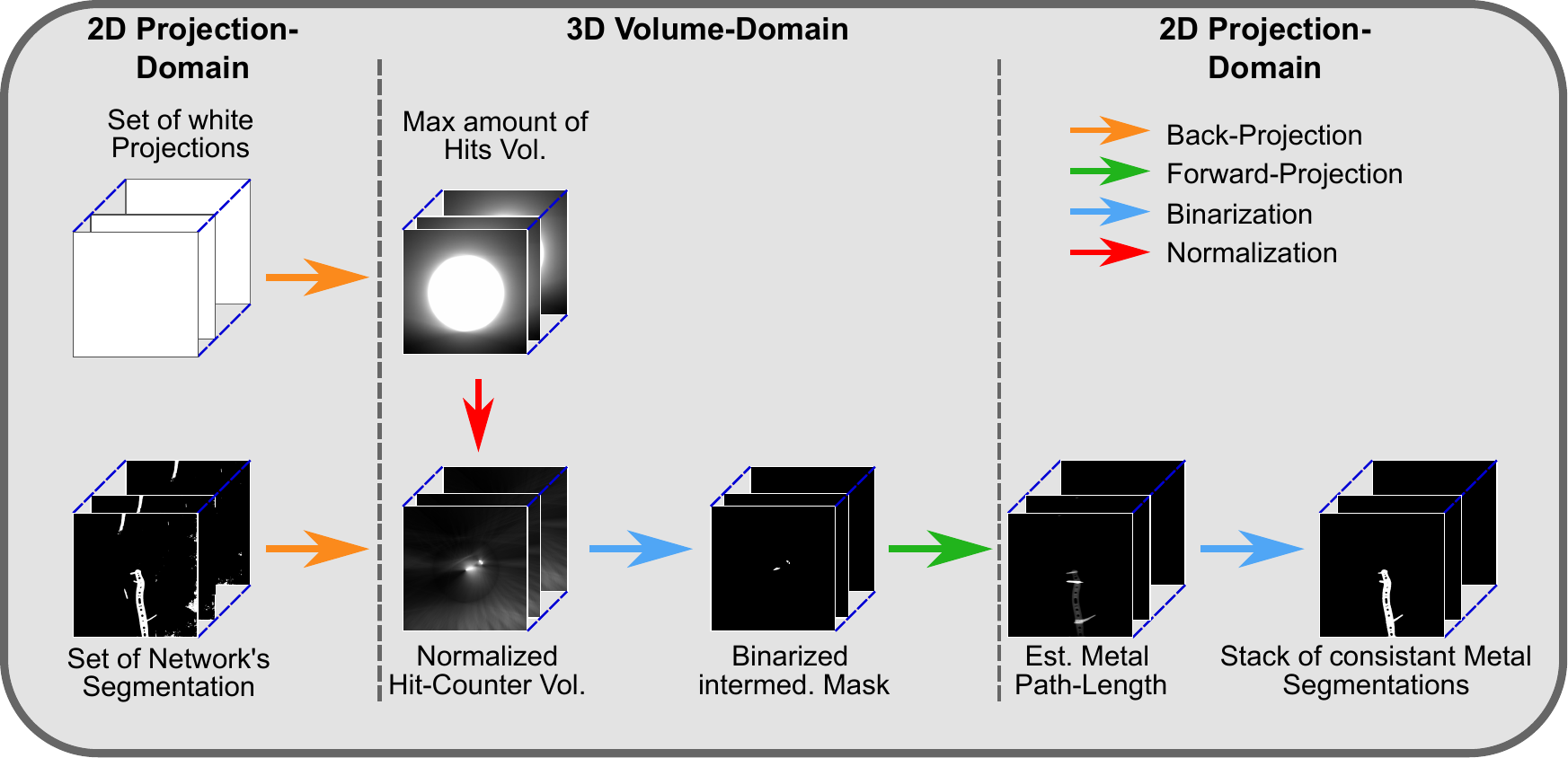}
\caption{Illustration of the single steps of the Consistency Filter post-processing.}
\label{fig:overview_CF}
\end{figure}

In \cite{Gottschalk2021}, we could already show the beneficial influence of the Consistency Filter as a post-processing step for the segmentation masks generated by the network. By exploiting the underlying consistency conditions of 3D reconstructions, the CF enforces consistency of the segmentation masks among the complete set of segmentation. Additionally, it can robustly remove false-positive segmentation of e.g. anatomical structures. An overview of the performed steps can be seen in Fig.~\ref{fig:overview_CF}.

Introducing the concept of the CF, we define three mappings \({M_{i}:\mathbb{R}^{2}\rightarrow\{0,1\}}\),\\ \({D_{i}:\mathbb{R}^{2}\rightarrow\{0,1\}}\) and \({P_{\text{i}}:\mathbb{R}^{3}\rightarrow\mathbb{R}^{2}}\). Whereas $P_{\text{i}}$ links a given 3D voxel $(x,y,z)$ with its corresponding 2D pixel $(l,m)$ of the current segmentation mask $i$,
\begin{equation}
    M_{i}(l,m)=
    \begin{cases}
        1, & \text{if  $(l,m)$ is segmented as metal in mask $i$} \\
        0, & \text{else}
    \end{cases}
\end{equation}
describes a function that checks whether a given pixel $(l,m)$ of the current mask $i$ is segmented as metal and
\begin{equation}
    D_{i}(l,m)=
    \begin{cases}
        1, & \text{if $(l,m)$ is inside of mask $i$} \\
        0, & \text{else}
    \end{cases}
\end{equation}
 describes a function that checks whether the given pixel $(l,m)$ is in general part of the mask $i$.
 
\begin{figure}[tb]
\centering
\includegraphics[scale=0.9]{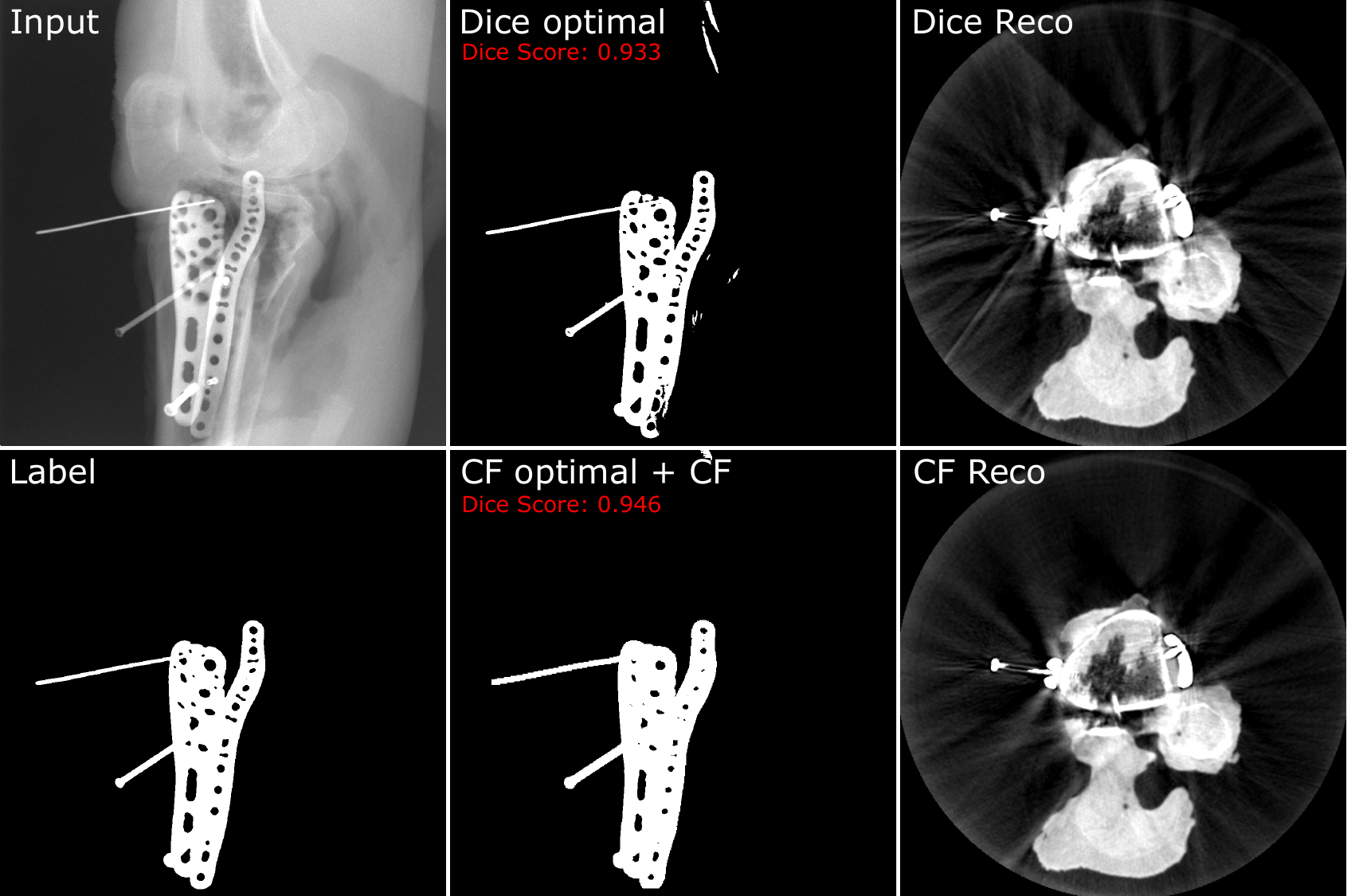}
\caption{Resulting segmentation masks for a given \textit{Input} projection. Whereas \textit{Label} shows the GT segmentation, \textit{Dice optimal} shows the network's binarized segmantation using a dice-optimal threshold, \textit{CF optimal + CF} shows the segmentation result after applying the Consitency Filter with an input segmentation using a CF-optimal threshold for binarization. Further, \textit{Dice Reco} and \textit{CF Reco} show the corresponding reconstructions. It can be seen, that in the case of the Dice-optimal reconstruction (thus not applying the CF post-processing), the lack of consistency and the false-positive segmentation lead to heavier streak artifacts. Additionally, the segmentation mask after applying the CF using a CF-optimal threshold achieves the higher Dice score.}
\label{fig:seg_results_CF}
\end{figure}

\noindent Further we define
\begin{equation} 
    V_{\text{Hit}}(x,y,z) = \displaystyle\sum_{i=1}^{N} M_{i}\left(P_{i}(x,y,z)\right)
\end{equation}
and
\begin{equation}
    V_{\text{Max}}(x,y,z) = \displaystyle\sum_{i=1}^{N} D_{i}\left(P_{i}(x,y,z)\right)\text{,}
\end{equation}
where $V_{\text{Hit}}$ can be understood as a ``hit-counter'', counting how many of the set of 2D segmentation masks contribute to that specific voxel and $V_{\text{Max}}$ expressing how many of the given segmentation masks can actually contribute. Further, $N$ denotes the number of segmentation masks, which is in our case 400.
Consequently, the normalization $V_{\text{Norm}}$ expressed by
 \begin{equation}
    V_{\text{Norm}}(x,y,z) = \frac{V_{\text{Hit}}(x,y,z)}{V_{\text{Max}}(x,y,z)}
\end{equation}
accounts for the decreasing number of possible hits towards the border of the reconstruction. Assuming that all pixels, mapped by $P_{\text{i}}$, are part of the current segmentation mask $i$, the normalized voxel values $V_{\text{Norm}}$ lie in the range of 0 and 1. Whereas voxels with a value of 0 correspond to no contribution of none of the segmentation masks, voxels with a value of 1 correspond to a contribution from every segmentation mask. As a consequence, the voxel values intrinsically hold information about how consistently the respective part of the metal object was segmented throughout the set of given projections. Thus, applying a threshold of e.g. 0.96 to binarize the normalized hit-counter volume to a intermediate 3D metal mask, is equivalent to including only those parts into the 3D metal mask, which were segmented in at least 96\% of the projections. Further, it is important to understand that this intermediate 3D mask only describes an overestimated envelope, because we back-projected binary 2D masks without any information about the thickness of the metal at each position. 
When subsequently applying the final forward-projection to that binarized 3D volume, consequently only consistently segmented metal parts are included in the final set of 2D segmentation masks. As a result, inconsistencies per se and thus also false-positive segmentations are excluded and also false-negatives are included up until a certain degree. Consequently, a clean and simultaneously consistent set of segmentation masks is provided. An example of the CF's beneficial influence on the segmentation result can be seen in Fig.~\ref{fig:seg_results_CF}. In the shown example, the segmentation network's output is once binarized using a threshold that optimizes the segmentation result concerning the Dice metric and once with a threshold that is optimized for the subsequently applied CF post-processing. The CF-optimized 2D segmentation shows no false-positives and filled-up false-negatives. Furthermore, the post-processed masks achieve the higher Dice score, although not being optimized for that metric. Moreover, the corresponding reconstructions show a higher amount of streak artifacts in the case of the Dice-optimal segmentation, which are mainly due to inconsistencies withing the set of segmentation.

Additionally, we want to clarify that the size of the normalized volume is chosen larger than the initial diagnostic reconstruction. Instead of 512$^3$ voxels with a voxel size of 0.313\,\textrm{mm}, it consists of 1500x1500x600 voxels with the same voxel size. This is because all present metal objects of the measure projections should be part of the normalized hit-counter volume. Thus, also metal objects that lie at the border of the projection image and hence outside the FoV of the initial reconstruction can still be parts of the final CF-processed segmentation masks.

\subsection{Experimental Setup}
\label{sec:exp_setup}

\subsubsection{Influence of the Patch-Size}
\label{sec:patch}
To evaluate the influence of different patch-sizes during the training phase as well as during inference, the described metal segmentation network was trained several times using patches of the sizes $64^2$ pixels, $128^2$ pixels, $256^2$ pixels, and $512^2$ pixels. Additionally, the network was trained one more time as an unpatched version, thus receiving the whole projection image of size $976^2$ pixels. Other than for the remaining experiments of this paper, the study regarding the patch-sizes was conducted extracting patches from the data pool as it is described in \cite{Gottschalk2021}. 
Despite its patch-based training and due to its fully-convolutional architecture, the proposed network was however applied with an unpatched inference strategy, thus receiving the complete projection image as input. Since the network's outputs are not yet binary segmentation masks, the outputs are binarized using a threshold, which was heuristically chosen to $0$.

\subsubsection{Influence of Proposed Method on the fsMAR performance}
\label{sec:ex_setup_reco}
In the interest of evaluating the influence of replacing the 3D volume-based thresholding mechanism of the standard fsMAR with the proposed 2D-projection-based data-driven metal segmentation, coupled with the CF, an evaluation on the 2D segmentation result itself does not suffice. This is because such an evaluation cannot give insights into the final MAR's performance. The influence of the proposed approach has to be evaluated on the final MAR's reconstructed volumes themselves. For this purpose, a five-fold cross-validation was performed, creating four different reconstructions for each test dataset within each of the folds. These reconstructions are shown in Fig.~\ref{fig:res_1} and Fig.~\ref{fig:res_2} and are denoted the following: The \textit{metal-free} and thus artifact-free GT volume is denoted as \textit{Label}. It was generated by reconstructing the acquired metal-free projection images coupled with a subsequently performed metal injection based on the fsMAR's 3D mask. The unprocessed reference volume which contains metal artifacts due to being reconstructed using the acquired metal-corrupted projection images is denoted as \textit{NoMAR}. The MAR volume that was reconstructed using the standard fsMAR method is denoted as \textit{fsMAR} and the modified fsMAR volume, which used the standard fsMAR steps except that our data-driven metal segmentation mechanism is applied, is denoted as \textit{Ours}. In the scope of this paper, the evaluation was limited to the results achieved using the outputs of the network trained with the patch-size of 256\textsuperscript{2} coupled with the unpatched inference and the proposed CF as post-processing. Whereas the threshold of binarizing the network's output was chosen to $0$, the threshold for the amount of demanded consistency enforced by the CF, was chosen to $96\%$ based on experiments as described in Section \ref{sec:Infl_CF}.

\begin{figure}[tbp]
\centering
\includegraphics[scale=0.8]{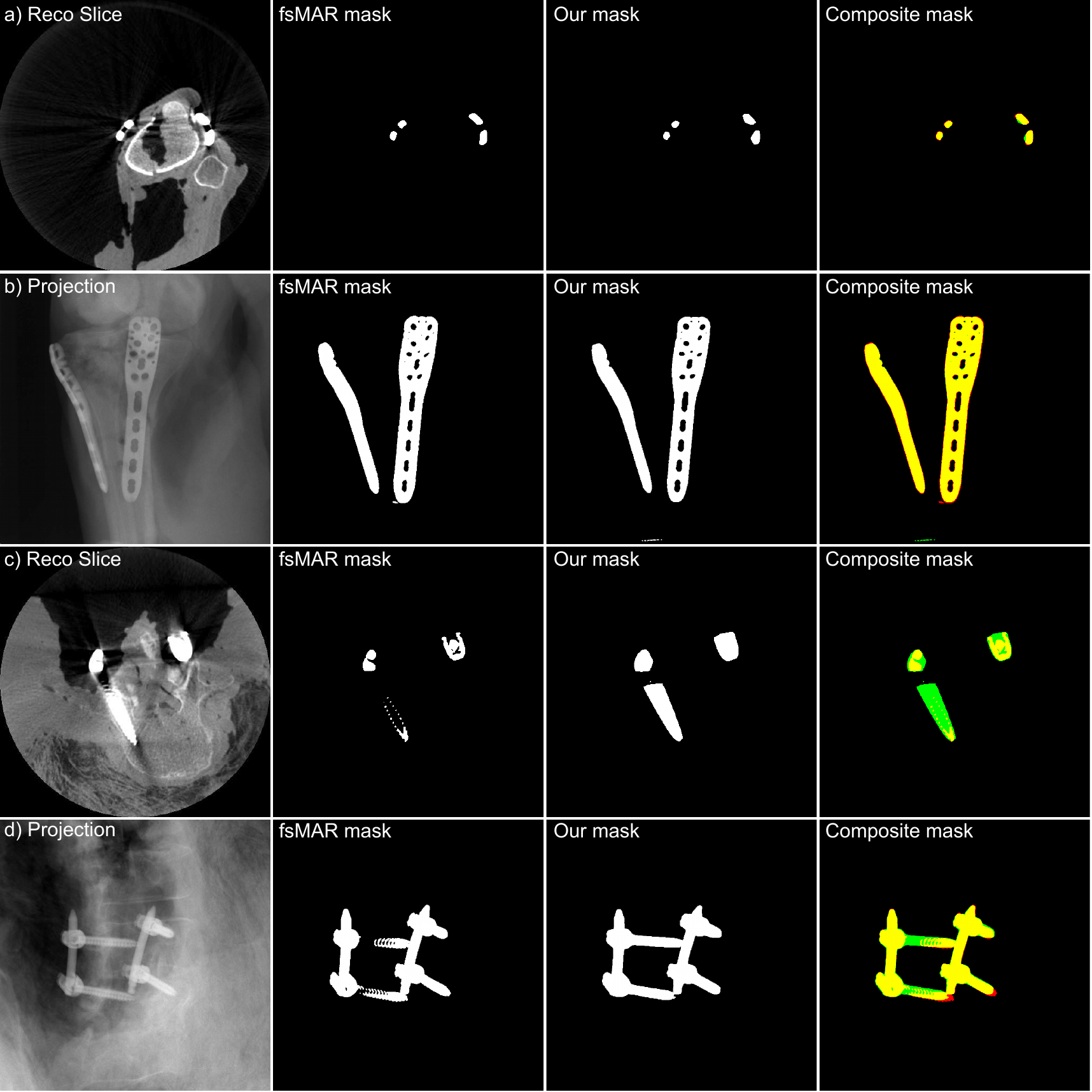}
\caption{The 1\textsuperscript{st} column shows two examples, each shown as a 3D reconstruction slice (cf. a) and c)) and corresponding 2D projection (cf. b) and d)), where a) and b) correspond to the first example (knee) and c) and d) to the second example (spine). Whereas the 2\textsuperscript{nd} column shows the threshold-based 3D masks and its corresponding forward-projected 2D mask extracted by the standard fsMAR, the 3\textsuperscript{rd} column represents the respective proposed modified fsMAR masks. Moreover, the 4\textsuperscript{th} row shows the composite/joint segmentation mask of both methods, which is used to mask the quantitative evaluation metrics as explained in Sec.~\ref{sec:ex_setup_reco}. It can be seen, that the masks are widely comparable for the first example, whereas the masks for the second example vary drastically. This is due to the CBCT typical artifacts and the inability to segment the resulting vanishing metal structures in a 3D segmentation approach.}
\label{fig:3d_seg_compare}
\end{figure}

To quantify the influence of the different segmentation mechanisms (standard fsMAR vs. modified fsMAR), the reduction of metal artifacts in the corresponding reconstructions was evaluated. This was done by calculating the Structural Similarity Index Measure (SSIM) \citep{Wang2004} as well as the Peak Signal to Noise Ratio (PSNR) between the \textit{Label} reconstruction and the different MAR methods reconstructions. Since the methods were evaluated using the acquired cadaver datasets, no 3D ground truth metal segmentation masks exist. This is because the 2D label segmentations, that are generated by the difference between the metal-free and metal-corrupted 2D projections, can only provide an overestimated envelope of the metal mask in 3D as already explained in Sec.~\ref{sec:CF}. Consequently, there is no possibility to create a real label reconstruction that is free of metal artifacts and at the same time comprises the correct undisturbed metal information. This is only possible when creating the test datasets by simulation. To still provide a meaningful evaluation with our cadaver data, only non-metal regions of the reconstructed slices are considered for the calculation of the metrics. This is sufficient for the evaluation because the metal artifacts spread across the whole size of the reconstructed slices. Additionally, anatomical structures adjacent to the metal implants are influenced differently by the compared approaches. To neither favor one of the two methods in the evaluation, a joint 3D segmentation mask of both methods is used for masking the metrics. This means, that all reconstructed regions that are included in that joint mask are set to $0$ in all reconstructions, thus not contributing to the different metrics results. This joint mask $M_{\text{Joint}}$ is calculated as follows:
\begin{equation}
    M_{\text{Joint}} = M_{\text{fsMAR}} \vee M_{\text{mod.fsMAR}}\text{,}
\end{equation}
where $M_{\text{fsMAR}}$ denotes the binary 3D mask generated by the standard fsMAR and $M_{\text{mod.fsMAR}}$ the binarized intermediate 3D mask of the modified fsMAR (cf. Fig.~\ref{fig:overview_CF}). To make sure that all metal implants are safely included in that joint mask, the proposed segmentation approach is parametrized (solely for creating that joint mask; that parameter set is not applied during inference of the test datasets) that it creates slightly overestimated masks. Therefore the network's output threshold is set to $-5$ and the CF's threshold for the amount of necessary consistency is decreased to $95\%$. Thus, an overall slightly grown joint mask is achieved. Two examples of the different MAR's masks, as well as their joint mask, can be seen in Fig.~\ref{fig:3d_seg_compare}. Examples of excluded regions masked by the joint mask can e.g. be seen in Fig.~\ref{fig:res_1} where they are denoted with red outlines.

Apart from quantitatively evaluating the influence on the test datasets, the influence was also qualitatively evaluated on two clinical datasets to investigate the generalization abilities of the approach. The mentioned data was acquired by the BG Trauma Centre Ludwigshafen, Germany during real clinical interventions.

\subsubsection{Influence of the Consistency Filter Threshold} \label{sec:Infl_CF}
To investigate the influence of the threshold that enforces different degrees of consistency within the CF, as explained in Section 2.2.2, experiments are performed with the first of the five data folds. We therefore examine the changes for different CF thresholds (0.8, 0.85, 0.9, 0.95, 0.96, 0.97, 0.98, 0.99, 0.992, 0.994, 0.996, 0.998) directly on the corresponding CF-processed 2D segmentation masks. This is done by calculating the precision, recall, and F-Score. Further, the influence on the image quality of the corresponding downstream MAR reconstructions is measured by the SSIM. Based on the mentioned reasons in Section \ref{sec:ex_setup_reco} the SSIM is again only calculated in non-metal areas by masking the reconstructions with the joint metal mask.

\section{Experimental Results}
\label{sec:results}

\subsection{Influence of the Patch-Size}

\begin{figure}[tb]
\centering
\includegraphics[scale=1.0]{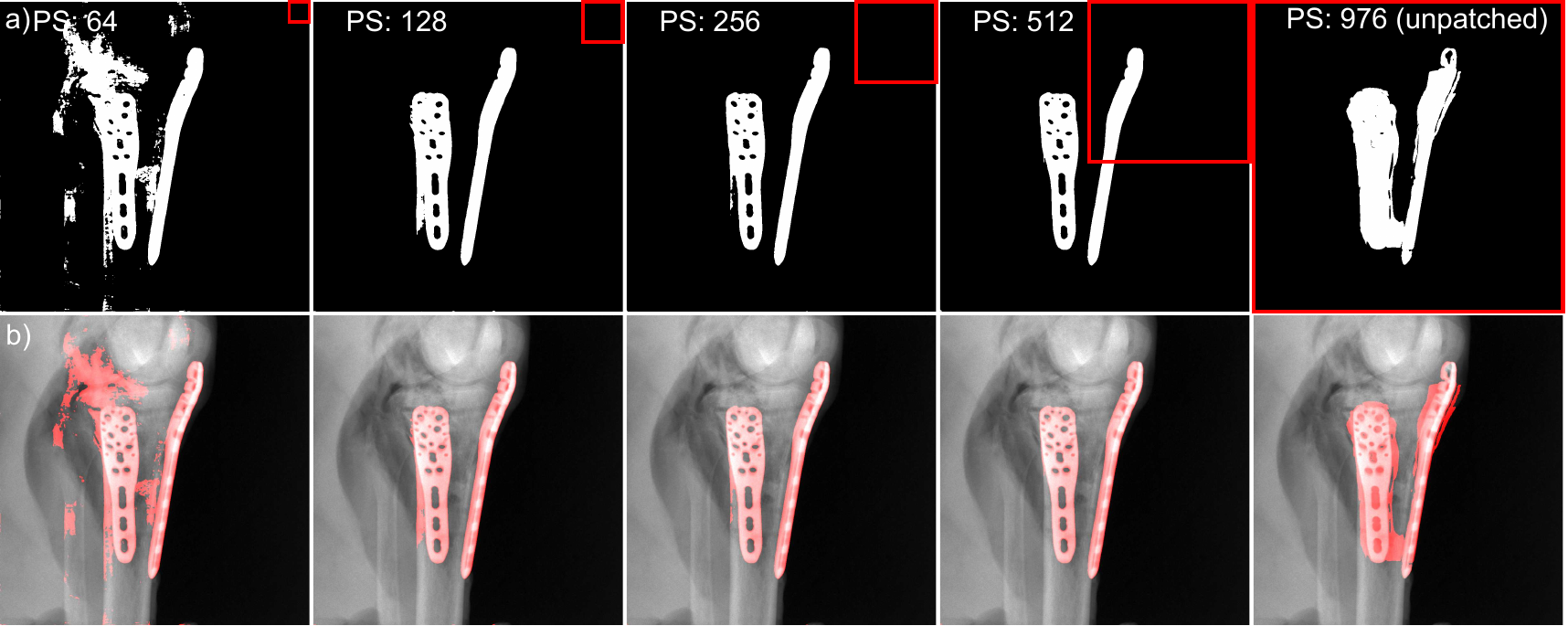}
\caption{The image shows the segmentation results of the different segmentation strategies. a) shows the network's results being trained with increasing patch-sizes (PS) from left to right. The respective patch-sizes are additionally illustrated as red squares with there actual sizes. b) shows the respective composite images, which overlay the corresponding segmentation result with the network's input projection.}
\label{fig:2d_seg_comp}
\end{figure}

The influence of the different patch-sizes during training and the unpatched inference was quantitatively evaluated using the receiver operating characteristic curves (RoC) and its corresponding area under the curve (AUC) values on pixel-level. These metrics are used because they are independent of the chosen threshold to binarize the network's output. The AUC can reach a maximum of 1. The overview about the micro-mean AUCs, where micro-mean denotes the mean AUC values over all 400 segmentations of one test scan, can be found in Tab.~\ref{tab:res_seg}. Additionally, segmentation results of one qualitative example are presented in Fig.~\ref{fig:2d_seg_comp}.

Investigating the quantitative evaluation in Tab.~\ref{tab:res_seg} it becomes apparent that using a patch-based training of the network leads to higher AUCs than training the network on the complete projection with a size of 976\textsuperscript{2}. This observation holds for all tested training patch-sizes with achieving the lowest macro-mean and macro-median AUC values (where \textit{macro} denotes the mean over the micro-means of all test scans) of 0.97343 and 0.97861, respectively, when using the unpatched trained network. Furthermore, it can be observed that the performance increases with increasing patch-size, reaching a plateau for patch-sizes 256 and 512, performing rather equivalently well. Whereas training with patch-size 512 reaches the highest macro-mean AUC value of 0.99784, patch-size 256 achieves the highest macro-median AUC value of 0.99951.

\begin{table}[bt]
\centering
\caption{Overview about the micro-mean AUC values for the different trained patch-sizes. Each line of the table represents the micro-mean AUC values over all 400 projection images of one test scan (TS). The last two rows show the macro-mean and macro-median AUC values over all TS, respectively. The best overall results, as well as the best results for each specific TS, are represented in bold type.}
\label{tab:res_seg}

\begin{tabular}{@{}llllll@{}}
\toprule
         & \multicolumn{5}{c}{Patch-Size}                                    \\ \cmidrule(lr){2-6}
Test Set & 64      & 128     & 256              & 512              & unpatched     \\ \midrule
TS1      & 0.99787 & 0.99922 & \textbf{0.99928} & 0.99926          & 0.99337 \\
TS2      & 0.98798 & 0.99982 & \textbf{0.99987} & 0.99985          & 0.99522 \\
TS3      & 0.97892 & 0.98248 & 0.98174          & \textbf{0.98617} & 0.97369 \\
TS4      & 0.99529 & 0.99985 & \textbf{0.99986} & 0.99966          & 0.99532 \\
TS5      & 0.98515 & 0.99962 & \textbf{0.99975} & 0.99952          & 0.98715 \\
TS6      & 0.98806 & 0.99952 & \textbf{0.99972} & 0.99945          & 0.99456 \\
TS7      & 0.99541 & 0.99856 & 0.99896          & \textbf{0.99926} & 0.97861 \\
TS8      & 0.98677 & 0.99301 & 0.99381          & \textbf{0.99425} & 0.95824 \\
TS9      & 0.98766 & 0.99962 & 0.99970          & \textbf{0.99972} & 0.98187 \\
TS10     & 0.99326 & 0.99936 & \textbf{0.99951} & 0.99948          & 0.97156 \\
TS11     & 0.99437 & 0.99955 & \textbf{0.99967} & 0.99964          & 0.96814 \\
TS12     & 0.98522 & 0.99757 & 0.99771          & \textbf{0.99781} & 0.96592 \\
TS13     & 0.99173 & 0.99827 & \textbf{0.99864} & 0.99787          & 0.89096 \\ \midrule
Mean     & 0.98982 & 0.99742 & 0.99756          & \textbf{0.99784} & 0.97343 \\
Median   & 0.98806 & 0.99936 & \textbf{0.99951} & 0.99945          & 0.97861 \\ \bottomrule
\end{tabular}
\end{table}

The observed trends can also be investigated in the presented example results qualitatively shown in Fig.~\ref{fig:2d_seg_comp}. The worst performance can be observed for the network trained on the complete projection, causing a high amount of false-positive segmentation and even a small amount of false-negative segmentation (cf. Fig.~\ref{fig:2d_seg_comp}; upper right corner). In contrast, all patch-based networks achieve a high amount of true-positive segmentation. However, the smallest patch-size of 64 generates a significant amount of false-positives. Furthermore, we see that an increasing patch-size decreases the amount of false-positive segmentation while keeping the true-positive segmentation rather constant.

\subsection{Influence of the Consistency Filter Threshold}

\begin{figure}[tb]
\centering
\includegraphics[scale=0.35]{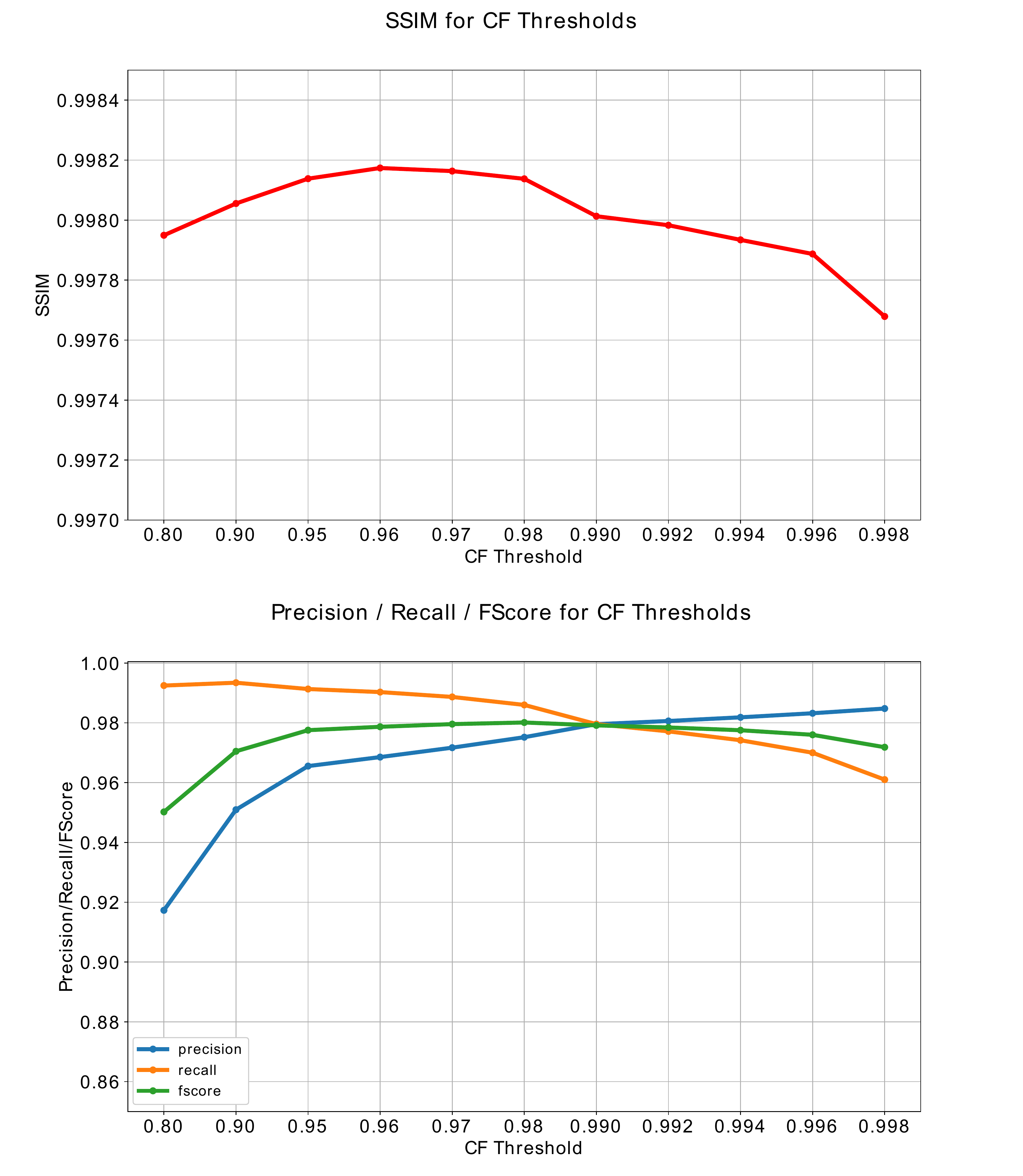}
\caption{
Overview about the SSIM and F-score, precision and recall scores for the different Consistency Filter thresholds.
}
\label{fig:ablation_CF}
\end{figure}

The line plots of the CF threshold experiments regarding the precision, recall, and F-score calculated on the CF-processed 2D segmentation masks themselves and the SSIM of the corresponding MAR reconstructions are presented in Fig.~\ref{fig:ablation_CF}. It becomes apparent that a rather low consistency requirement of 0.8 leads to low precision, F-Score, and SSIM scores. In contrast, the recall reaches its maximum. Further, an increasing enforced consistency causes increasing precision, F-Score, and SSIM, whilst the recall decreases. Between a threshold of 0.95 and 0.996, a rather moderate increase of the precision and decrease of the recall can be observed. The F-Score builds a plateau in this range. At the threshold of 0.990 precision and recall and consequently also the F-Score are crossing. The threshold of 0.998 creates a slightly higher decrease in recall and F-Score. Focusing on the SSIM and thus on the image quality of the corresponding reconstructions reveals a peak SSIM score of 0.9982 at a threshold of 0.96 with a subsequent decrease when applying increasing thresholds. Equivalent to the observed effects on the recall and F-Score a more significant drop-off of image quality can be seen for threshold 0.998.

\subsection{Influence of Proposed Method on the fsMAR Performance}

Since the proposed projection-based segmentation method is only one step of an inpainting-based MAR method, and an evaluation on the segmentation itself cannot give insights into the final MAR's performance, the influence of the proposed approach has to be evaluated on the reconstructed volumes. In the scope of this paper, this is done using the network that is trained with the patch-size of 256, coupled with an unpatched inference and the proposed CF as post-processing. This configuration was chosen over the equivalently performing network with patch-size 512 since it reaches the highest micro-mean AUCs in the maximal amount of test sets (denoted in bold type in Tab.~\ref{tab:res_seg}). Further, the CF threshold was set to 0.96 according to the performed experiments. Using this set of parameters, the network was trained and tested five times using the different data folds.

\begin{table}[tb]
\centering
\caption{Mean SSIM and PSNR values over all folds of the cross-validation. The last row presents the mean metrics' values over all folds. The best result for each fold and the best overall are shown in bold type.}
\label{tab:res_metrics_folds}
\begin{tabular}{lllllll}
\toprule
   & \multicolumn{3}{c}{SSIM}                  & \multicolumn{3}{c}{PSNR}                     \\ \cmidrule(lr){2-4} \cmidrule(lr){5-7}
Fold & NoMAR        & fsMAR        & Ours         & NoMAR         & fsMAR         & Ours          \\ \midrule
1  & 0.9939 & 0.9973 & \textbf{0.9981} & 60.06 & 62.40 & \textbf{63.12} \\
2  & 0.9954 & 0.9976 & \textbf{0.9978} & 61.41 & 63.20 & \textbf{63.38} \\
3  & 0.9950 & 0.9969 & \textbf{0.9980} & 61.07 & 62.54 & \textbf{63.15} \\
4  & 0.9938 & 0.9965 & \textbf{0.9979} & 60.67 & 62.40 & \textbf{63.08} \\
5  & 0.9957 & 0.9977 & \textbf{0.9983} & 61.26 & 62.97 & \textbf{63.43} \\ \midrule
\O & 0.9948 & 0.9972 & \textbf{0.9981} & 60.90 & 62.70 & \textbf{63.23} \\ \bottomrule
\end{tabular}
\end{table}

Before focusing on more detailed results for the first data fold, Tab.~\ref{tab:res_metrics_folds} shows the overview about the mean SSIM and PSNR scores over all folds. It can be investigated that the modified segmentation method slightly elevates the MAR performance leading to slightly increased overall SSIM and PSNR scores for all folds of the validation. The mean SSIM and PSNR score of the modified fsMAR lies $0.9e^{-3}$ and $0.5$ dB, respectively, above those of the standard fsMAR. Furthermore, both, the resulting SSIM and PSNR scores for the standard fsMAR as well as modified fsMAR, only vary marginally over the different folds. The maximum deviation of the SSIM of the standard fsMAR and modified fsMAR between the folds lie at $1.2e^{-3}$ and $0.5e^{-3}$, respectively. For the PSNR scores at $0.8$ and $0.3$ dB.

\begin{table}[tb]
\centering
\caption{Intra-fold Mean SSIM and PSNR values over all metal-containing slices of the masked volumes of the test scans (TS) contained in the first fold of the cross-validation. Furthermore, the maximal difference (Max Diff.) between corresponding slices of the standard fsMAR and modified FsMAR is shown. Whereas positive differences are in favor of the modified fsMAR, negative differences are in favor of the standard fsMAR. The last row presents the macro-mean metrics' values over all TS of the fold. The best result for each TS and the best overall (fold-mean) are shown in bold type.}
\label{tab:res_metrics}
\begin{tabular}{lllllllll}
\toprule
   & \multicolumn{4}{c}{SSIM}                  & \multicolumn{4}{c}{PSNR}                     \\ \cmidrule(lr){2-5} \cmidrule(lr){6-9}
TS & NoMAR        & fsMAR        & Ours   & Max Diff.       & NoMAR         & fsMAR         & Ours & Max Diff.          \\ \midrule
1  & 0.9958 & 0.9984 & \textbf{0.9986} & 0.0060 & 62.15 & 63.73 & \textbf{63.81} & 3.004\\
2  & 0.9935 & 0.9974 & \textbf{0.9983} & 0.0294 & 59.26 & 62.24 & \textbf{63.11} & 8.649\\
3  & 0.9957 & 0.9978 & \textbf{0.9981} & 0.0067 & 58.54 & 60.35 & \textbf{60.51} & 3.965\\
4  & 0.9955 & 0.9969 & \textbf{0.9971} & 0.0028 & 59.86 & 61.45 & \textbf{61.64} & 1.327\\
5  & 0.9944 & 0.9976 & \textbf{0.9980} & 0.0035 & 60.42 & 62.81 & \textbf{63.27} & 2.517\\
6  & 0.9910 & 0.9943 & \textbf{0.9971} & 0.0280 & 59.41 & 60.94 & \textbf{62.38} & 8.667\\
7  & 0.9945 & 0.9984 & \textbf{0.9994} & 0.0076 & 60.71 & 64.12 & \textbf{65.86} & 9.028\\
8  & 0.9953 & 0.9982 & \textbf{0.9985} & 0.0099 & 61.81 & 63.45 & \textbf{63.56} & 3.857\\
9  & 0.9943 & 0.9974 & \textbf{0.9980} & 0.0146 & 60.53 & 62.39 & \textbf{62.73} & 5.002\\
10 & 0.9932 & 0.9974 & \textbf{0.9982} & 0.0081 & 59.18 & 62.49 & \textbf{63.39} & 6.724\\
11 & 0.9894 & 0.9961 & \textbf{0.9981} & 0.0134 & 58.78 & 62.41 & \textbf{64.03} & 6.822\\ \midrule
\O & 0.9939 & 0.9973 & \textbf{0.9981} & 0.0118 & 60.06 & 62.40 & \textbf{63.12} & 5.415\\ \bottomrule
\end{tabular}
\end{table}

\begin{figure}[tb]
\centering
\includegraphics[scale=1.0]{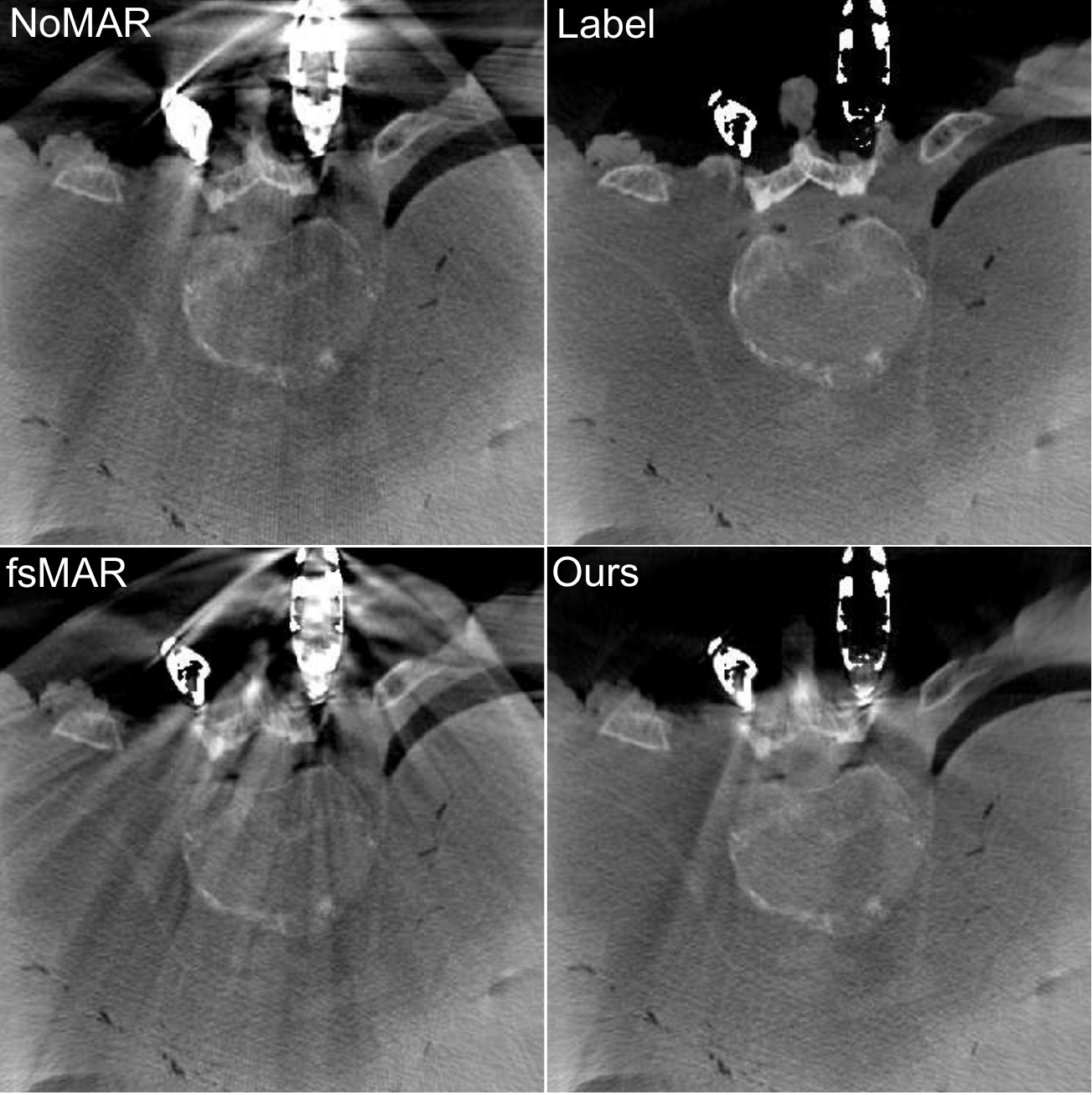}
\caption{Reconstruction results for TS 8 of fold 4, which comprises a high amount of metal objects. Additional to pedicle screws, this dataset also includes towers, which are used during minimal invasive spine surgery. Whereas the standard fsMAR can only slightly reduce the artifacts due to a failed segmentation mask, the modified fsMAR achieves a significant artifact reduction.}
\label{fig:res_towers}
\end{figure}

Investigating the results in more detail for the first fold of the cross-validation, the quantitative comparison between the standard fsMAR and the proposed modified fsMAR can be found in Tab.~\ref{tab:res_metrics}. It becomes apparent that the switch to the proposed segmentation approach increases the quality of the MAR's results, reaching a fold-mean SSIM score of 0.9981 and a fold-mean PSNR value of 63.177 dB. Thus, the modified fsMAR surpasses the results of both, the unprocessed reconstruction, as well as the standard fsMAR in all 11 test scans. The highest difference over all slices shows for test scan 7 with an increase of 1.74 dB concerning the PSNR. Moreover, the maximal difference between the slices of the standard fsMAR and the modified fsMAR show more significant results. With a mean maximal difference of 0.0118 and 5.415 dB w.r.t. SSIM and PSNR respectively, the modified fsMAR clearly outperforms the standard fsMAR in certain slices. Further evaluations on other folds of the cross-validation show that increases regarding the mean over all slices with up to roughly 3 dB can be reached for more complex cases. These cases comprise a higher amount and more challenging metal configurations. Such an example is illustrated in Fig.~\ref{fig:res_towers}.

\begin{figure}[p]
\centering
\includegraphics[scale=0.85]{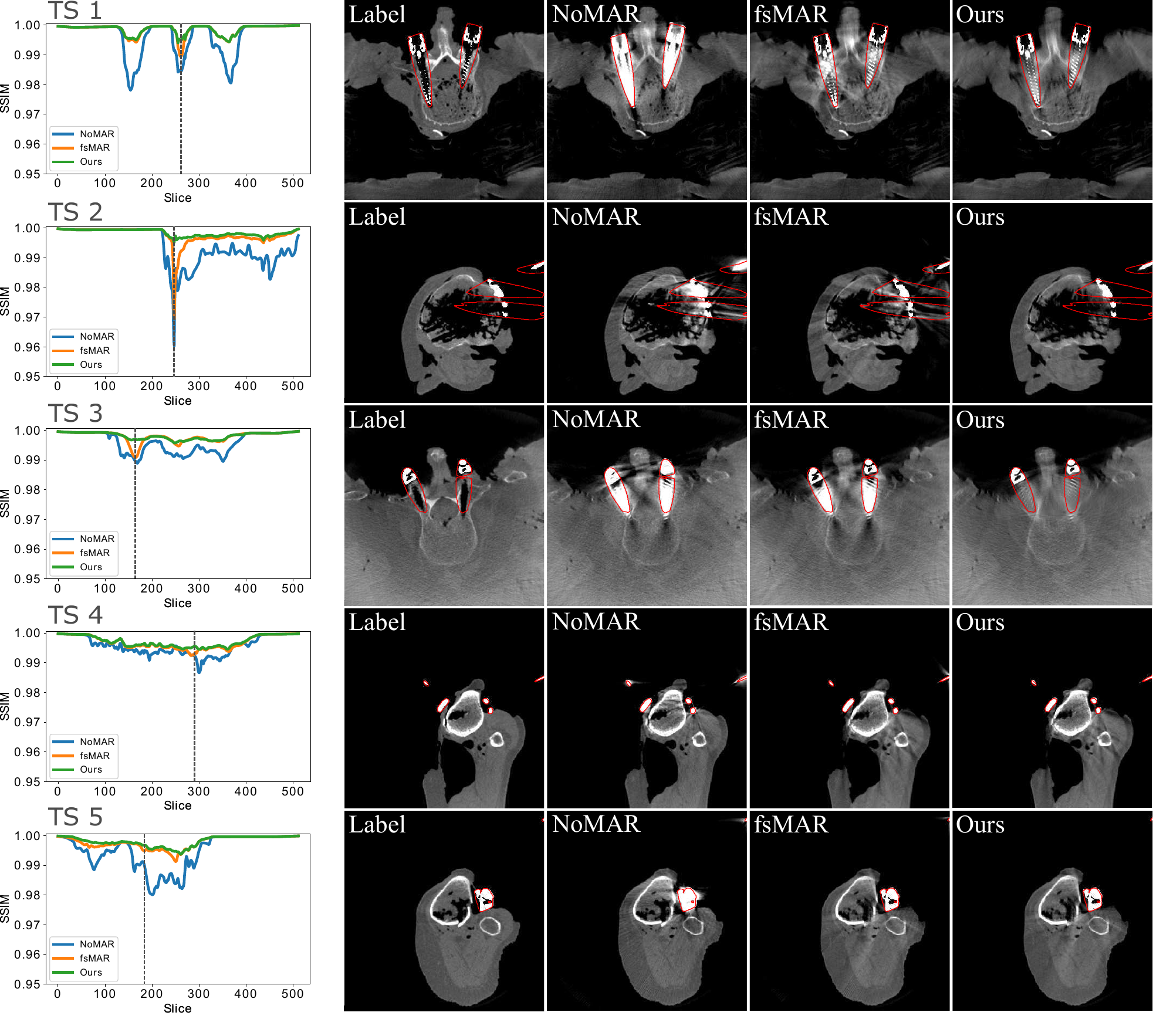}
\caption{Overview about the per-slice SSIM scores and one slice (marked as a gray dotted line) of each of the corresponding reconstructions for the first 5 TS of the first fold. Each row comprises the quantitative (left) and qualitative (right) results of one test scan. The 1\textsuperscript{st} column of the qualitative results shows the label reconstructions, the 2\textsuperscript{nd} column shows the reconstructions without the use of a MAR method (NoMAR), the 3\textsuperscript{rd} column shows the reconstruction using the standard fsMAR method and the 4\textsuperscript{th} column shows the corresponding reconstruction of using our proposed segmentation method for the modified fsMAR (Ours). The red outlines represent the envelope of the joint segmentation masks which are used to mask the metrics' scores. All regions inside the denoted envelope are set to $0$ during the metric calculations. All shown reconstructions are windowed with [500, 2048] HU.}
\label{fig:res_1}
\end{figure}

\begin{figure}[tbp]
\centering
\includegraphics[scale=0.85]{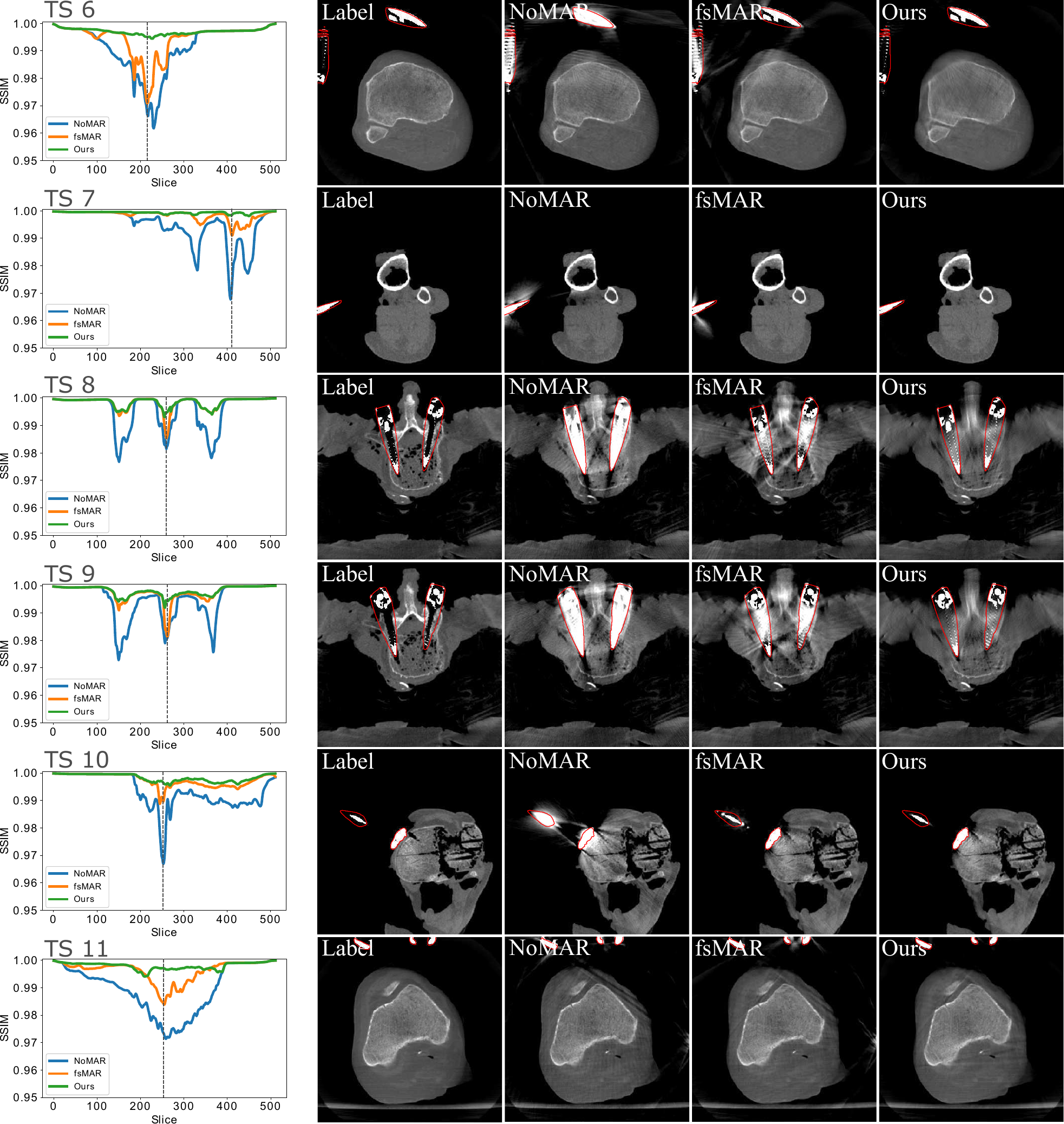}
\caption{Overview about the per-slice SSIM scores and one slice (marked as a gray dotted line) of each of the corresponding reconstructions for the last 6 TS of the first fold. The layout and annotations are equivalent to those of Fig.~\ref{fig:res_1}.}
\label{fig:res_2}
\end{figure}

By examining the qualitative results and the corresponding SSIM line plots presented in Fig.~\ref{fig:res_1} illustrating the results for test scans 1 to 5 and Fig.~\ref{fig:res_2} illustrating the results for test scans 6 to 11 of the first fold, similar effects can be observed. It becomes apparent that for the majority of test cases the standard fsMAR, as well as the modified fsMAR, perform equally well, with a slight beneficial tendency for the proposed approach. The differences are that marginal, that they do not visually show up in the corresponding presented slices (cf. Fig.~\ref{fig:res_1} \& Fig.~\ref{fig:res_2}). This especially holds for scans comprising knees. However, investigating the qualitative results for test scans 2, 3 (Fig.~\ref{fig:res_1}), 6, 8, 9 (Fig.~\ref{fig:res_2}) and the example in Fig.~\ref{fig:res_towers}, significant differences can be observed. In Fig.~\ref{fig:res_1} at test scan 3 comprising a spine, it can be seen that the standard fsMAR is not able to segment the complete inserted pedicle screws, but only the associated heads. Due to that, only artifacts originating from the screws' heads are reduced, whereas present beam hardening artifacts at the tips of the screws remain unchanged. In opposition to that, the proposed approach shows reduced artifacts throughout the complete size of the screws. Similar effects can be observed for test scans 8 and 9 in Fig.~\ref{fig:res_2} where the segmentation of the standard fsMAR misses parts of the screws. For test scan 2 (Fig.~\ref{fig:res_1}) it becomes apparent that the streak artifacts, which originate from parts of K-wires, which lie outside the FoV of the reconstruction, are reduced more efficiently using the proposed segmentation approach. The represented slice for test scan 6 (Fig.~\ref{fig:res_2}) shows a similar case in which metal reaches from outside the FoV into the reconstruction. Whereas the standard fsMAR is able to fully segment the present metal plate and only an insufficient part of the metal lying outside and at the border of the volume, the modified fsMAR segments all present metal parts, thus achieving a more artifact reduced reconstruction. For both cases, a significant performance difference can also be seen in the corresponding line plots. However, similar to the quantitative results, the most significant difference between the two approaches can be seen for the presented test scan in Fig.~\ref{fig:res_towers}. While for the shown slice, the standard fsMAR is not able to reduce any of the artifacts due to a failed segmentation, our approach nearly completely removes the present artifacts. Furthermore, the superior performance is not only reached in the single presented slice but rather holds for the complete set of slices.

\begin{figure}[tb]
\centering
\includegraphics[scale=0.85]{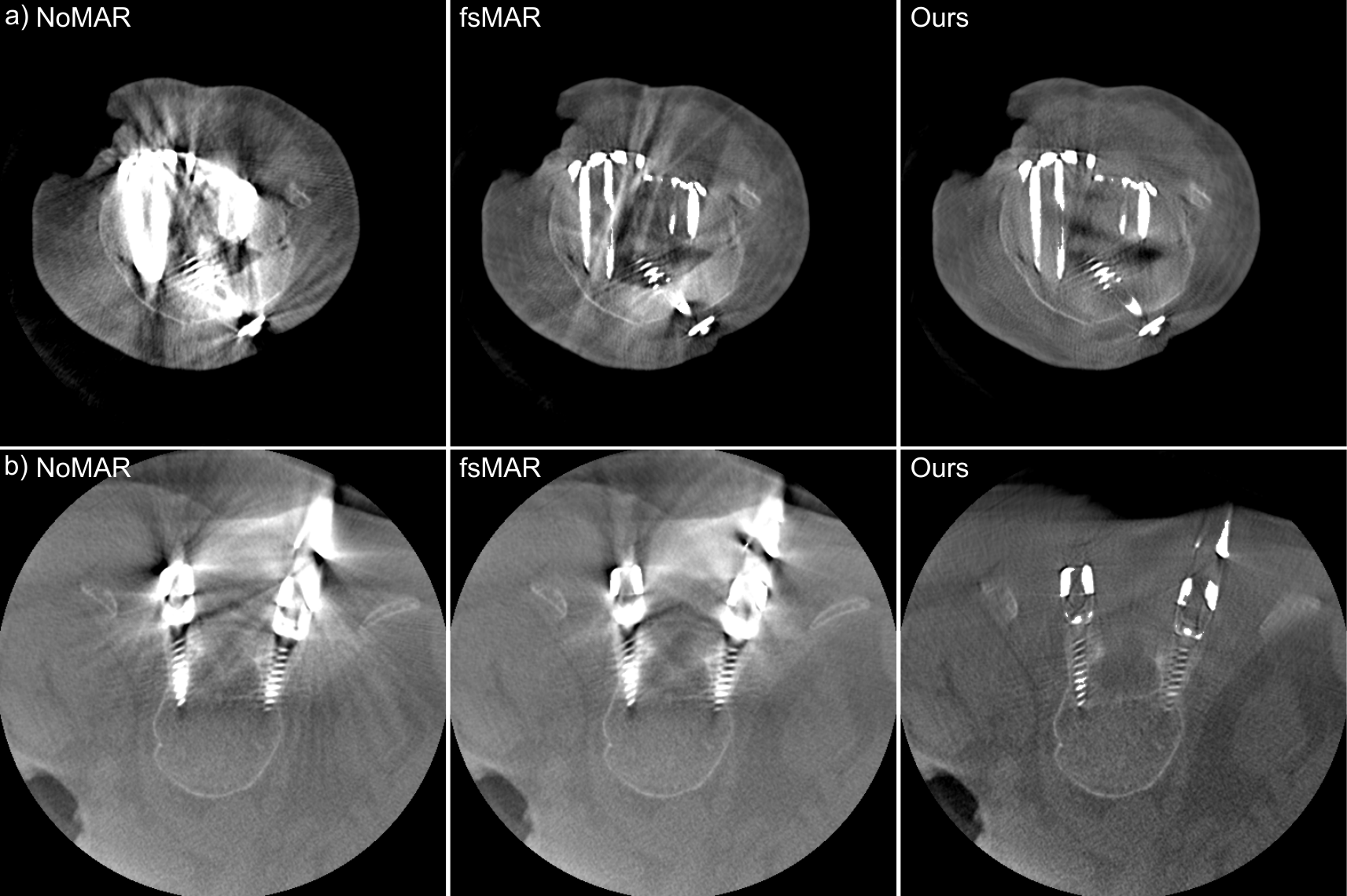}
\caption{Comparison of qualitative results for two clinical datasets. Row a) shows the MAR's reconstructions for a clinical knee dataset and row b) for a clinical spine dataset.}
\label{fig:clinical_res}
\end{figure}

When additionally examining the qualitative results for two clinical datasets, the same effects can be observed (cf. Fig.~\ref{fig:clinical_res}). Vanishing intensities in the metal objects lead to an under-segmentation of the implants by the standard fsMAR, thus reaching only a minor reduction of artifacts. The proposed approach however can segment the corresponding metal implants in 2D, thus reaching a significantly higher reduction of artifacts in both shown cases.

\section{Discussion}
\label{sec:discussion}
The presented results show that in cases that contain slices that comprise a severe artifacts e.g. originating from metal outside the FoV or vanishing metal intensities due to scatter and photon starvation, the shift of the segmentation to the 2D projection-domain coupled with the proposed CF is beneficial. Since the metal-containing slices commonly comprise the crucial information for an intraoperative evaluation of fracture reduction and correct implant positioning, the modified segmentation approach holds great potential for orthopedic/trauma interventions.

To discuss the results in more detail, we start by assessing the quantitative and qualitative results of training the network with different patch-sizes in combination with an unpatched inference strategy. It becomes apparent that patch-based training is beneficial for the segmentation task. This is probably due to the higher amounts of variation in the training dataset, caused by a higher amount of augmentation. Furthermore, the results reveal that the unpatched inference strategy achieves satisfying results, although the networks were trained patch-based. Performing an unpatched inference additionally holds the benefit of an increased inference speed compared to a patch-based approach using a sliding window with overlapping patches. Since the training data for the segmentation was mainly acquired for human knees and spine in orthopedic/trauma intervention setups, we do not necessarily expect the method to generalize to all conceivable metal objects like e.g. coils or stents. We assume that the method requires retraining or refinement for those objects.

The shown results of the experiments regarding the CF threshold give interesting insights into the behavior of the CF. The low threshold of $0.8$ results in a majorly overestimated intermediate 3D mask, thus comprising almost all true-positives (TP) but also a high amount of false-positives (FP). Whereas the high TPs explain the high recall, the high FPs result in a significant drop-off of the precision. The mentioned overestimation is due to including streak artifacts, which themselves originate from inconsistencies within the network's predicted masks, into the intermediate 3D mask. This happens when requiring a too low amount of consistency for the final segmentation masks. Despite the mentioned overestimation of the masks and the clear drop-off of the precision, the corresponding SSIM score is relatively high. This is because the broad masks lead to a rather large inpainting region, resulting in an almost artifact-free, but also smoothed reconstruction. However, in the context of surgical interventions, this is critical, because that smoothing could lead to blurring or even to the disappearance of important adjacent anatomical structures to the implant. Consequently, it is important to find a trade-off between removing the artifacts requiring a big enough mask to include all TPs, and retaining all important adjacent structures. This however requires a mask as small as possible to keep the amount of FPs low.
The significant increase of the precision between $0.8$ and $0.95$ is due to excluding the majority of artifacts from the intermediate 3D mask resulting in a shrinking of the mask. Furthermore, this comes only with a small loss of TPs, as can be seen by the rather stable recall and the increasing F-Score. The plateauing F-Score for the thresholds in the interval $0.95$ to $0.996$ further suggests a balanced increase and decrease of the TP and FP, respectively. The SSIM however shows a decrease starting with a threshold $0.99$ after almost plateauing between $0.95$ and $0.98$. The shrinking mask leads to losing TPs, thus leaving metal regions unprocessed by the downstream MAR, ending up with still present metal artifacts in the reconstruction. This behavior leads to a decreased SSIM and aggravates with increasing thresholds. Concluding the mentioned effects and the combination of the presented F-Score and SSIM suggests picking a CF threshold between $0.95$ and $0.98$ to fully leverage the benefits of the proposed CF. Due to the highest reached SSIM score at $0.96$, this threshold is conducted for the performed cross-validation.

When evaluating the results of the influence of shifting the segmentation task to 2D, the only slightly varying results over the different folds of the cross-validation show that there is no selection bias or overfitting. Since the presented mean scores of the cross-validation comprise averaging over all metal-containing slices of each single test scan, and additionally over all test scans of a fold, the results are somewhat diluted. In the majority of slices, neither effects like photon starvation nor metal reaching outside the FoV are present. Consequently, the standard fsMAR's threshold-based 3D segmentation works properly, hence, reaching similar metric scores as the modified fsMAR.

However, focusing on the detailed results of the first fold, we see several interesting effects. As explained, for simple cases in which the metal objects neither lie outside the FoV nor cause vanishing intensities inside the very same, the standard fsMAR, and the modified fsMAR show nearly equal quantitative results. The evaluation suggests only a slight advantage towards our proposed method. This is e.g. the case for test sets 1, 4, 5, 7, and 10. The corresponding qualitative results support that investigation showing no significant visual differences. However, for cases that comprise either metal outside the FoV or metal parts that suffer from a vanished representation in 3D (cf. test scans 2, 3, 6, 8, 9 and 11), clear advantages can be investigated applying the proposed 2D projection-based segmentation. The standard fsMAR's segmentation is unable to segment such metal parts, thus leaving those unprocessed in the downstream inpainting step. This leads to still clearly visible artifacts. In contrast, the results suggest that the modified segmentation still provides complete masks, thus enabling the downstream processing to be applied to all present metals. This yields a significantly higher amount of reduced artifacts. The very same impact appears even more prominent in the test scan presented in  Fig.~\ref{fig:res_towers} for TS 8 of fold 4, which comprises both effects -- metal outside the FoV and vanishing metal parts due to e.g. photon starvation. The same holds for the shown results for clinical data.

When exclusively looking at the inter-fold (Tab.~\ref{tab:res_metrics_folds}) and fold (Tab.~\ref{tab:res_metrics}) mean scores, it gives the impression that shifting the segmentation only marginally improves the results. However, during surgical interventions, not the overall image quality of a certain 3D scan (expressed by mean over all slices) is the main focus, but rather the quality of single slices that are crucial for the correct evaluation of e.g. the positioning of an implant. An undiluted difference of the both methods in such slices is better expressed by the presented maximal difference in Tab.~\ref{tab:res_metrics}. Therefore, the mean maximal difference of 5.415 dB reveals a clear advantage of the modified fsMAR in those slices and consequently a potential benefit for surgical applications.

Another interesting insight is revealed by the qualitative results for test sets with photon starvation effects (cf. e.g. Fig.~\ref{fig:res_1}, TS 3) reveal an interesting insight. As shown in Fig.~\ref{fig:basic_mar}, the inpainting step of the MAR uses a 2D metal mask, whereas the metal insertion step still requires a 3D metal mask. Thus, in the case of the standard fsMAR, these masks ``match'' because the 2D mask was acquired by forward-projection of the respective 3D mask. However, when applying our modified fsMAR, which uses our data-driven segmented 2D masks for the downstream inpainting while retaining the original fsMAR's 3D mask for the metal insertion, discrepancies between the masks occur. This has the effect of seemingly disappearing metal implants in cases of e.g. vanishing metal intensities, e.g. due to photon starvation. The more accurate 2D masks of our approach lead to regions being inpainted that are subsequently not being inserted back into the volume, due to being missed in the original 3D mask. Thus, further improvements need to include a new 3D segmentation approach or a mechanism that provides a consistent joint mask.

Since the proposed segmentation network, coupled with the CF was trained using real data from cadaver studies, we are convinced that the shown results generalize well to real clinical cases of surgical interventions. This assumption is further supported by the qualitative results shown for the two tested clinical cases, which reveal an increased MAR performance, using our proposed method (cf. Fig.~\ref{fig:clinical_res}). Moreover, the majority of the more recent learning-based MAR method relies on a 3D volume-based segmentation, thus suffering from the same inability to cope with those mentioned effects. Consequently, we believe that the proposed method can equally boost their performances for cases of e.g. metal outside the FoV. However, this should be proven with further experiments.

\section{Conclusion}
\label{sec:conclusion}

In this paper, we investigated the influence of shifting the segmentation task from a 3D volume-based thresholding to a data-driven, view-consistent, and 2D projection-wise approach for the MAR's downstream inpainting task. While the projection-wise segmentation is performed by a network trained on real cadaver data, the view-consistency of the segmentation is ensured by a subsequent CF. Experiments regarding the patch-size suggest coupling a patch-based training with an unpatched inference of the segmentation network. Experiments w.r.t. the proposed segmentation method's influence on the downstream MAR task, reveal that the proposed approach has a strongly beneficial influence on the amount of reduced artifacts in crucial slices -- especially in cases that comprise metal objects outside the FoV, or an vanished representation of the very same. Further, the shifted segmentation is intrinsically invulnerable to the metal-artifacts themselves.
However, the inconsistency between the proposed 2D mask for the inpainting task and the 3D mask for the metal injection leads to disappearing metal implants and thus insufficient image quality. Since the proposed method was trained on cadaver data with (at least partly) clinical-like realism, the method is expected to generalize well for real surgical interventions. This assumption is supported by the qualitative results shown for the two clinical cases. Furthermore, we are convinced that the shown improvements are reproducible with all MAR methods currently relying on a 3D thresholding-based metal segmentation coupled with a subsequent 2D-based inpainting step.

Summarizing the conducted experiments shows that the proposed shift of the segmentation to a view-consistent 2D projection-domain-based approach achieves equivalent results to a 3D volume-based thresholding when being applied to cases with moderate amounts of artifacts. However, being applied to cases comprising severe artifacts that lead to vanishing metal intensities, the proposed shift holds significant benefits.


\acks{The authors gratefully acknowledge funding of the Erlangen Graduate School in Advanced Optical Technologies (SAOT) by the Bavarian State Ministry for Science and Art. Furthermore, the research leading to these results has received funding from the European Research Council (ERC) under the European Union’s Horizon 2020 research and innovation program (ERC grant no. 810316). Additionally, the authors like to thank Siemens Healthineers for their extensive support of this project and the Rimasys GmbH for their help during the cadaver studies.}

%
\ethics{The work follows appropriate ethical standards in conducting research and writing the manuscript, following all applicable laws and regulations regarding treatment of animals or human subjects, or cadavers of both kind. All data acquisitions were done in consultation with the Institutional Review Board of the University Hospital of Erlangen, Germany.
}

\coi{We declare that T.M.~Gottschalk, F.~Kordon and B.W.~Kreher do not have a conflict of interest. A.~Maier is Associate Editor of the Journal of Machine Learning for Biomedical Imaging (MELBA).} \\

\dis{The methods and information presented here are based on research and are not commercially available.}


\bibliography{references}

\begin{thebibliography}{27}
\providecommand{\natexlab}[1]{#1}
\providecommand{\url}[1]{\texttt{#1}}
\expandafter\ifx\csname urlstyle\endcsname\relax
  \providecommand{\doi}[1]{doi: #1}\else
  \providecommand{\doi}{doi: \begingroup \urlstyle{rm}\Url}\fi

\bibitem[Abadi et~al.(2016)Abadi, Barham, Chen, Chen, Davis, Dean, Devin,
  Ghemawat, Irving, Isard, et~al.]{abadi2016}
Mart{\'\i}n Abadi, Paul Barham, Jianmin Chen, Zhifeng Chen, Andy Davis, Jeffrey
  Dean, Matthieu Devin, Sanjay Ghemawat, Geoffrey Irving, Michael Isard, et~al.
\newblock Tensorflow: A system for large-scale machine learning.
\newblock In \emph{12th $\{$USENIX$\}$ symposium on operating systems design
  and implementation)}, pages 265--283, 2016.

\bibitem[Badal and Badano(2009)]{Badal2009}
Andreu Badal and Aldo Badano.
\newblock Accelerating monte carlo simulations of photon transport in a
  voxelized geometry using a massively parallel graphics processing unit.
\newblock \emph{Med. Phys.}, 36\penalty0 (11):\penalty0 4878--4880, 2009.
\newblock \doi{https://doi.org/10.1118/1.3231824}.
\newblock URL
  \url{https://aapm.onlinelibrary.wiley.com/doi/abs/10.1118/1.3231824}.

\bibitem[Badrinarayanan et~al.(2017)Badrinarayanan, Kendall, and
  Cipolla]{badrinarayanan2017}
Vijay Badrinarayanan, Alex Kendall, and Roberto Cipolla.
\newblock Segnet: A deep convolutional encoder-decoder architecture for image
  segmentation.
\newblock \emph{IEEE Trans Pattern Anal Mach Intell}, 39\penalty0
  (12):\penalty0 2481--2495, 2017.

\bibitem[Barrett and Keat(2004)]{barrett2004}
Julia~F. Barrett and Nicholas Keat.
\newblock Artifacts in ct: recognition and avoidance.
\newblock \emph{Radiographics}, 24\penalty0 (6):\penalty0 1679--1691, 2004.

\bibitem[Chen et~al.(2018)Chen, Papandreou, Kokkinos, Murphy, and
  Yuille]{Chen2018}
Liang-Chieh Chen, George Papandreou, Iasonas Kokkinos, Kevin Murphy, and
  Alan~L. Yuille.
\newblock Deeplab: Semantic image segmentation with deep convolutional nets,
  atrous convolution, and fully connected crfs.
\newblock \emph{IEEE Transactions on Pattern Analysis and Machine
  Intelligence}, 40\penalty0 (4):\penalty0 834--848, 2018.
\newblock \doi{10.1109/TPAMI.2017.2699184}.

\bibitem[Claus et~al.(2017)Claus, Jin, Gjesteby, Wang, and De~Man]{Claus2017}
Bernhard~E.H. Claus, Yannan Jin, Lars~A. Gjesteby, Ge~Wang, and Bruno De~Man.
\newblock {Metal-artifact reduction using deep-learning based sinogram
  completion: Initial results}.
\newblock In \emph{Proc. 14th Int. Meeting Fully Three-Dimensional Image
  Reconstruction Radiol. Nucl. Med.}, pages 631--634, 2017.

\bibitem[Drozdzal et~al.(2016)Drozdzal, Vorontsov, Chartrand, Kadoury, and
  Pal]{Drozdzal2016}
Michal Drozdzal, Eugene Vorontsov, Gabriel Chartrand, Samuel Kadoury, and Chris
  Pal.
\newblock The importance of skip connections in biomedical image segmentation.
\newblock In Gustavo Carneiro, Diana Mateus, Lo{\"i}c Peter, Andrew Bradley,
  Jo{\~a}o Manuel R.~S. Tavares, Vasileios Belagiannis, Jo{\~a}o~Paulo Papa,
  Jacinto~C. Nascimento, Marco Loog, Zhi Lu, Jaime~S. Cardoso, and Julien
  Cornebise, editors, \emph{Deep Learning and Data Labeling for Medical
  Applications}, pages 179--187, Cham, 2016. Springer International Publishing.
\newblock ISBN 978-3-319-46976-8.
\newblock URL \url{https://doi.org/10.1007/978-3-319-46976-8\_19}.

\bibitem[Ghani and Karl(2018)]{Ghani2018}
Muhammad~Usman Ghani and W.~Clem Karl.
\newblock {Deep Learning Based Sinogram Correction for Metal Artifact
  Reduction}.
\newblock \emph{Electron. Imaging}, 2018\penalty0 (15):\penalty0 472--1--4728,
  2018.
\newblock URL \url{https://doi.org/10.2352/ISSN.2470-1173.2018.15.COIMG-472}.

\bibitem[Gjesteby et~al.(2017)Gjesteby, Yang, Xi, Shan, Claus, Jin, Man, and
  Wang]{Gjesteby2017}
Lars Gjesteby, Qingsong Yang, Yan Xi, Hongming Shan, Bernhard Claus, Yannan
  Jin, Bruno~De Man, and Ge~Wang.
\newblock {Deep learning methods for CT image-domain metal artifact reduction}.
\newblock In Bert Müller and Ge~Wang, editors, \emph{Developments in X-Ray
  Tomography XI}, volume 10391, pages 147 -- 152. International Society for
  Optics and Photonics, SPIE, 2017.

\bibitem[Gottschalk et~al.(2021)Gottschalk, Kreher, Kordon, and
  Maier]{Gottschalk2021}
Tristan~M. Gottschalk, Bj{\"o}rn~W. Kreher, Florian Kordon, and Andreas Maier.
\newblock Learning-based patch-wise metal segmentation with consistency check.
\newblock In \emph{BVM 2021}, 2021.

\bibitem[{He} et~al.(2015){He}, {Zhang}, {Ren}, and {Sun}]{He2015}
Kaiming {He}, X.~{Zhang}, S.~{Ren}, and J.~{Sun}.
\newblock Delving deep into rectifiers: Surpassing human-level performance on
  imagenet classification.
\newblock In \emph{2015 IEEE International Conference on Computer Vision},
  pages 1026--1034, 2015.

\bibitem[Huang et~al.(2018)Huang, Wang, Tang, Zhong, and Zhang]{Huang2018}
Xia Huang, Jian Wang, Fan Tang, Tao Zhong, and Yu~Zhang.
\newblock {Metal artifact reduction on cervical CT images by deep residual
  learning}.
\newblock \emph{Biomedical engineering online}, 17\penalty0 (1):\penalty0 175,
  2018.
\newblock URL \url{https://doi.org/10.1186/s12938-018-0609-y}.

\bibitem[Isensee et~al.(2019)Isensee, Kickingereder, Wick, Bendszus, and
  Maier-Hein]{Isensee2019}
Fabian Isensee, Philipp Kickingereder, Wolfgang Wick, Martin Bendszus, and
  Klaus~H. Maier-Hein.
\newblock No new-net.
\newblock In Alessandro Crimi, Spyridon Bakas, Hugo Kuijf, Farahani Keyvan,
  Mauricio Reyes, and Theo van Walsum, editors, \emph{Brainlesion: Glioma,
  Multiple Sclerosis, Stroke and Traumatic Brain Injuries}, pages 234--244,
  Cham, 2019. Springer International Publishing.
\newblock ISBN 978-3-030-11726-9.
\newblock URL \url{https://doi.org/10.1007/978-3-030-11726-9\_21}.

\bibitem[Isensee et~al.(2021)Isensee, Jaeger, Kohl, Petersen, and
  Maier-Hein]{isensee2021}
Fabian Isensee, Paul~F Jaeger, Simon~AA Kohl, Jens Petersen, and Klaus~H
  Maier-Hein.
\newblock nnu-net: a self-configuring method for deep learning-based biomedical
  image segmentation.
\newblock \emph{Nat. Methods}, 18\penalty0 (2):\penalty0 203--211, 2021.

\bibitem[Ketcha et~al.(2021)Ketcha, Marrama, Souza, Uneri, Wu, Zhang, Helm, and
  Siewerdsen]{ketcha2021}
Michael~D Ketcha, Michael Marrama, Andre Souza, Ali Uneri, Pengwei Wu, Xiaoxuan
  Zhang, Patrick~A Helm, and Jeffrey~H Siewerdsen.
\newblock Sinogram+ image domain neural network approach for metal artifact
  reduction in low-dose cone-beam computed tomography.
\newblock \emph{Journal of Medical Imaging}, 8\penalty0 (5):\penalty0 052103,
  2021.

\bibitem[Kingma and Ba(2014)]{kingma2014}
Diederik~P. Kingma and Jimmy Ba.
\newblock {Adam: A Method for Stochastic Optimization}, 2014.

\bibitem[Long et~al.(2015)Long, Shelhamer, and Darrell]{Long2015}
Jonathan Long, Evan Shelhamer, and Trevor Darrell.
\newblock Fully convolutional networks for semantic segmentation.
\newblock In \emph{Proceedings of the IEEE Conference on Computer Vision and
  Pattern Recognition}, June 2015.

\bibitem[Maier et~al.(2018)Maier, Steidl, Christlein, and Hornegger]{maier2018}
Andreas Maier, Stefan Steidl, Vincent Christlein, and Joachim Hornegger.
\newblock \emph{{Medical Imaging Systems: An Introductory Guide}}, volume
  11111.
\newblock Springer, 2018.

\bibitem[Meyer et~al.(2010)Meyer, Raupach, Lell, Schmidt, and
  Kachelriess]{Meyer2010}
Esther Meyer, Rainer Raupach, Michael Lell, Bernhard Schmidt, and Marc
  Kachelriess.
\newblock {Normalized metal artifact reduction (NMAR) in computed tomography}.
\newblock \emph{Med. Phys.}, 37\penalty0 (10):\penalty0 5482--5493, 2010.

\bibitem[Meyer et~al.(2012)Meyer, Raupach, Lell, Schmidt, and
  Kachelrieß]{Meyer2012}
Esther Meyer, Rainer Raupach, Michael Lell, Bernhard Schmidt, and Marc
  Kachelrieß.
\newblock {Frequency split metal artifact reduction (FSMAR) in computed
  tomography}.
\newblock \emph{Med. Phys.}, 39\penalty0 (4):\penalty0 1904--1916, 2012.
\newblock URL \url{https://doi.org/10.1118/1.3691902}.

\bibitem[Park et~al.(2018)Park, Lee, Kim, Seo, and Chung]{Park2018}
Hyoung~Suk Park, Sung~Min Lee, Hwa~Pyung Kim, Jin~Keun Seo, and Yong~Eun Chung.
\newblock {CT sinogram-consistency learning for metal-induced beam hardening
  correction}.
\newblock \emph{Med. Phys.}, 45\penalty0 (12):\penalty0 5376--5384, 2018.
\newblock URL \url{https://doi.org/10.1002/mp.13199}.

\bibitem[Ronneberger et~al.(2015)Ronneberger, Fischer, and
  Brox]{Ronneberger2015}
Olaf Ronneberger, Philipp Fischer, and Thomas Brox.
\newblock {U-Net: Convolutional Networks for Biomedical Image Segmentation}.
\newblock In Nassir Navab, Joachim Hornegger, William~M. Wells, and
  Alejandro~F. Frangi, editors, \emph{Medical Image Computing and
  Computer-Assisted Intervention -- MICCAI 2015}, pages 234--241, Cham, 2015.
  Springer International Publishing.

\bibitem[Stille et~al.(2013)Stille, Kratz, Müller, Maass, Schasiepen, Elter,
  Weyers, and Buzug]{Stille2013}
Maik Stille, Bärbel Kratz, Jan Müller, Nicole Maass, Ingo Schasiepen,
  Matthias Elter, Imke Weyers, and Thorsten~M. Buzug.
\newblock {Influence of Metal Segmentation on the Quality of Metal Artifact
  Reduction Methods}.
\newblock In \emph{Medical Imaging 2013: Physics of Medical Imaging}, volume
  8668, pages 902 -- 907. International Society for Optics and Photonics, SPIE,
  2013.
\newblock URL \url{https://doi.org/10.1117/12.2006810}.

\bibitem[Unberath et~al.()Unberath, Zaech, Lee, Bier, Fotouhi, Armand, and
  Navab]{Unberath2018}
Mathias Unberath, Jan-Nico Zaech, Sing~Chun Lee, Bastian Bier, Javad Fotouhi,
  Mehran Armand, and Nassir Navab.
\newblock {DeepDRR--A Catalyst for Machine Learning in Fluoroscopy-guided
  Procedures}.
\newblock In \emph{Proc. Medical Image Computing and Computer Assisted
  Intervention}. Springer.

\bibitem[Xinhui et~al.(2008)Xinhui, Li, Yongshun, Jianping, Zhiqiang, and
  Yuxiang]{Xinhui2008}
Duan Xinhui, Zhang Li, Xiao Yongshun, Cheng Jianping, Chen Zhiqiang, and Xing
  Yuxiang.
\newblock {Metal artifact reduction in CT images by sinogram TV inpainting}.
\newblock In \emph{2008 IEEE Nucl. Sci. Symp. Conf. Rec.}, pages 4175--4177, 10
  2008.

\bibitem[Yu et~al.(2021)Yu, Zhang, Li, and Xing]{Yu2021}
Lequan Yu, Zhicheng Zhang, Xiaomeng Li, and Lei Xing.
\newblock Deep sinogram completion with image prior for metal artifact
  reduction in ct images.
\newblock \emph{IEEE Transactions on Medical Imaging}, 40\penalty0
  (1):\penalty0 228--238, 2021.
\newblock \doi{10.1109/TMI.2020.3025064}.

\bibitem[{Zhou Wang} et~al.(2004){Zhou Wang}, {Bovik}, {Sheikh}, and
  {Simoncelli}]{Wang2004}
{Zhou Wang}, A.~C. {Bovik}, H.~R. {Sheikh}, and E.~P. {Simoncelli}.
\newblock Image quality assessment: from error visibility to structural
  similarity.
\newblock \emph{{IEEE Transactions on Image Processing}}, 13\penalty0
  (4):\penalty0 600--612, 4 2004.

\end{thebibliography}

\appendix 

\noindent


\end{document}